\documentclass[aps,groupaddress,superscriptaddress,amssymb,onecolumn]{revtex4-2}

\usepackage[utf8]{inputenc}
\usepackage{color}
\usepackage{xcolor}
\usepackage{graphicx}

\usepackage{mdframed}
\usepackage{physics}
\usepackage{amsmath}
\usepackage{amsfonts}
\usepackage{bm}
\usepackage[symbol]{footmisc}

\usepackage{lineno}

\begin{document}
\preprint{APS/123-QED}

\title{The emergence of scale-free fires in Australia}

\author{Giorgio Nicoletti}
\affiliation{Laboratory of Interdisciplinary Physics, Department of Physics and Astronomy ``G. Galilei'', University of Padova, Padova, Italy}
\affiliation{Department of Mathematics ``T. Levi-Civita'', University of Padova, Padova, Italy}
\author{Leonardo Saravia}
\affiliation{Centro Austral de Investigaciones Científicas (CADIC - CONICET), Ushuaia, Argentina}
\affiliation{Área Biología y Bioinformática, Universidad Nacional de General Sarmiento, Buenos Aires, Argentina}
\author{Fernando Momo}
\affiliation{Universidad Nacional de General Sarmiento, Instituto de Ciencias, Área Biología y Bioinformática, Buenos Aires, Argentina}
\affiliation{Instituto de Ecología y Desarrollo Sustentable (INEDES - CONICET), Buenos Aires, Argentina}
\author{Amos Maritan}
\affiliation{Laboratory of Interdisciplinary Physics, Department of Physics and Astronomy ``G. Galilei'', University of Padova, Padova, Italy}
\author{Samir Suweis}
\affiliation{Laboratory of Interdisciplinary Physics, Department of Physics and Astronomy ``G. Galilei'', University of Padova, Padova, Italy}

\begin{abstract}
Between 2019 and 2020, during the country’s hottest and driest year on record, Australia experienced a dramatic bushfire season, with catastrophic ecological and environmental consequences. Several studies highlighted how such abrupt changes in fire regimes may have been in large part a consequence of climate change and other anthropogenic transformations. Here, we analyze the monthly evolution of the burned area in Australia from 2000 to 2020, obtained via satellite imaging through the MODIS platform. We find that the 2019-2020 peak is associated with signatures typically found near critical points. We introduce a modeling framework based on forest-fire models to study the properties of these emergent fire outbreaks, showing that the behavior observed during the 2019-2020 fire season matches the one of a percolation transition, where system-size outbreaks appear. Our model also highlights the existence of an absorbing phase transition that might be eventually crossed, after which the vegetation cannot recover.
\end{abstract}

\maketitle

\section*{Introduction}
\noindent Bushfires are an intrinsic part of Australia's landscape dynamics. Its natural ecosystems have evolved to coexist with fires, and mitigation strategies to reduce their impact have been learned in the most vulnerable areas \cite{yates2008big}. Yet, the 2019-2020 fire season was particularly catastrophic. It began in July 2019, at the end of the country’s hottest and driest year on record, and wildfires were unprecedented in their spatial extent and severity \cite{lindenmayer2020new, deb2020causes, ward2020impact, phillips2020race}. In the eastern Australia states of New South Wales and Victoria, around 5.8 million hectares of mainly temperate broadleaf forest were burned by a series of high-impact fires, many of which exceeded a size of $100,000 \mathrm{ha}$ and continued to burn for weeks after ignition. Several studies highlighted how this abrupt departure from the historical trend may have been in large part a consequence of climate change and other anthropogenic transformations \cite{dowdy2018climatological, phillips2020race, yu2020bushfires, boer2020unprecedented, arriagada2020climate, abram2021connections}. Furthermore, these high-impact fires had a devastating effect on Australia's biodiversity. Of more than $830$ taxa - comprising birds, reptiles, frogs, mammals, and freshwater fish - around one-fourth lost to the fires between $10\%$ and $50\%$ of their Australian habitat, sixteen of them lost between $50\%$ and $80\%$, and three more than $80\%$ \cite{ward2020impact}.

These drastic changes, with their catastrophic effects on vegetation and on biodiversity, are often associated with critical transitions, i.e., conditions that inevitably lead to large-scale fire outbreaks and subsequent widespread damage \cite{scheffer2009early, van2016you, rocha2018cascading, scheffer2020critical}. Such behavior has been observed in many different systems ranging from Amazon forests \cite{wuyts2017amazonian, lovejoy2018amazon} to Kalahari vegetation \cite{scanlon2007positive} and more in general in tropical forests fragmentation \cite{taubert2018global, Saravia2018a}. In physical systems with many degrees of freedom, these phenomena are well-known to appear at the edge of phase transitions. When a system undergoes a continuous phase transition at a critical point, scale-free behaviors described by power-laws are found - such as long-range correlations and diverging susceptibility to external perturbations - due to the underlying scale-invariance that emerges at criticality \cite{binney1992theory, caldarelli2001, sornette2006critical, mckenzie2012power}. This lack of a characteristic scale is a possible mechanism behind the abrupt appearance of large and out-of-scale events, such as the high-impact fires experienced by Australia between 2019 and 2020.

In this work, we analyze the monthly evolution of the burned area in the East and Southeast temperate broadleaf and mixed forests of continental Australia \cite{Dinerstein2017}. These data, spanning from November 2000 to June 2020, are obtained via satellite imaging through the MODIS platform \cite{Giglio2016}, and allow us to analyze the spatiotemporal properties of fire propagation. Unsurprisingly, we find that the 2019-2020 peak of the burned area exceeds the historical data. Then, thanks to the high spatial resolution of the data, we study the distributions of spatially-separated clusters of burned area, as well as their evolution in time. By applying tools from Statistical Physics, we find that during 2019-2020 the distribution of fire outbreak sizes is compatible with a power-law, and it is invariant under spatial coarse-graining. Our results suggest that such fires lacked a characteristic size, and thus the system may have been poised at a critical point of their spreading dynamics \cite{sornette2006critical}.

To understand the drivers and the type of such critical transitions, we introduce a paradigmatic spatial model that describes the concurrent spreading of fires and vegetation over a two-dimensional lattice. In a regime where the timescale of fire propagation is much faster than the vegetation one, our numerical simulations suggest that the model predicts the crossing of a percolation-like transition \cite{stauffer2018introduction} to a more arid climate, where spreading becomes easier for the fires and harder for the vegetation. Differently from self-organized forest-fires models \cite{grassberger1991forestfires, drossel1992self, grassberger1993forestfires, Christensen1993forestfires, malamud1998forest, Turcotte2002forestfires, palmieri2020forest}, the dynamics of our model depends only on two effective ecological parameters. When these parameters cross the percolation-like critical point, the features of the model - such as the distribution of the fires' sizes - are qualitatively comparable to the ones observed during the bushfire season of 2019-2020 in Australia. This suggests that this kind of phase transition in the vegetation-fires dynamics may have been at the heart of the emergence of scale-free fire outbreaks. Our paradigmatic model encompasses another kind of critical point as well, that corresponds to an absorbing phase transition \cite{dickman1999, henkel2009} after which the vegetation cannot recover. Although possibly unrelated from an ecological perspective, it foreshadows how critical points may lead to further abrupt and fundamental changes in the fire-vegetation dynamics.

\section*{Results}
\subsection*{Fire sizes distributions}
\noindent In order to shed a light on how the 2019-2020 bushfire season emerged, we analyze the timeseries of the burned area obtained from $236$ monthly satellite images of Australia, spanning from November 2000 to June 2020 (Methods and Figure S1). Each month is represented by a binary matrix $M_t$, whose entries $(M_t)_{ij}$ represent an area of $500 \rm{m}^2$ and are set to $1$ if the corresponding pixel matches an area that has burned in the span of that month.

\begin{figure*}[t]
\centering
\includegraphics[width = 1\textwidth]{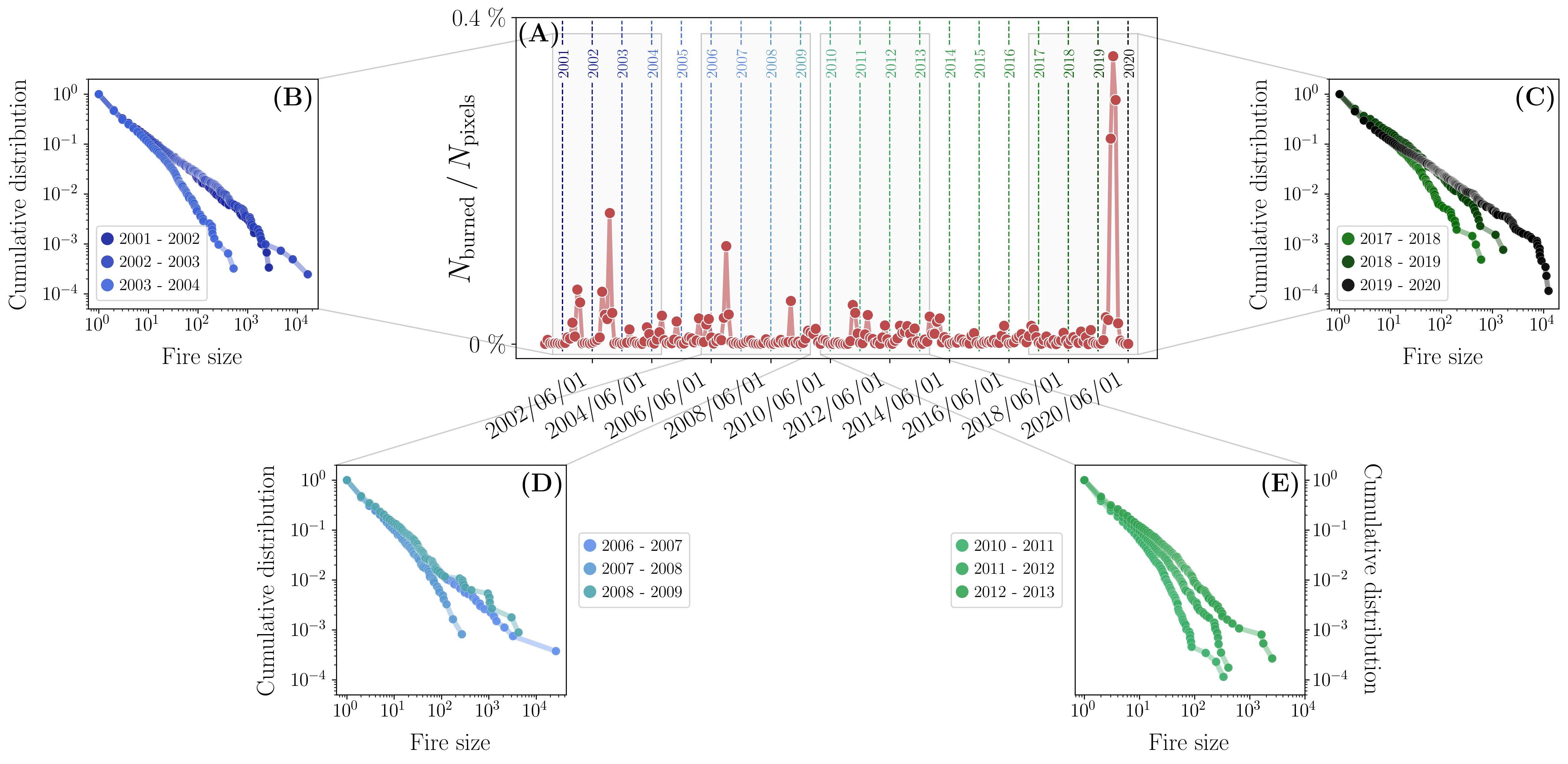}
\caption{\textbf{The cumulative distribution of the fire sizes at different years.} (A) The timeseries of the normalized number of burned pixels per month in the East and Southeast temperate broadleaf and mixed forests of continental Australia, $N_{\rm burned}(t)$, from 2000 to 2020, normalized by the total number of pixels. Years are defined as the twelve months occurring between June and May. The year 2019-2020 largely exceeds the peaks of the previous twenty years. (B-E) For a given year, we can compute the cumulative fire size distribution on a nearest-neighbor basis. Even though peaks, such as during 2002-2003, often display either longer tails in the distribution or are dominated by few, very large fires, a distinctive power-law behavior emerges during 2019-2020. See also Figure S2.}
\label{fig:timeseries_properties}
\end{figure*}

The exceptional nature of the 2019-2020 events is perhaps already striking from the timeseries of the total burned area spanning the last 20 years, $N_{\rm burned}(t) = \sum_{ij} (M_t)_{ij}$, plotted in Figure~\ref{fig:timeseries_properties}a. As a gauge of the extent of the damage, less than half of the pixels were burning during the second largest peak - which took place in the season 2002-2003. Most importantly, given the spatial nature of our data, we can also compute the cumulative distribution of clusters of burned pixels for a given month - on a nearest-neighbors basis - starting from $M_t$. Hence, for each month and each matrix $M_t$, we compute the number of clusters $n_c(t)$ and their sizes $\{A_c^{(i)}\}_{i=1}^{n_c(t)}$. We identify $n_c$ with the number of separate fire outbreaks, and $A_c^{(i)}$ with the corresponding outbreak sizes. In particular, we can compute the corresponding cumulative fire sizes distribution during a given year (Methods), defined from June to May to include the summer of the Southern Hemisphere.

These distributions, shown in Figure~\ref{fig:timeseries_properties}b-e, typically display longer tails during peaks of the burned area $N_{\rm burned}(t)$, whereas the sizes are exponentially suppressed if the overall burned area is low (see also Figure S2). In particular, higher peaks of $N_{\rm burned}(t)$ - e.g., the 2002-2003 or the 2019-2020 season - are associated with distributions that span a wide range of sizes. Although it is tempting to relate such distributions to power-laws, the finite size of the system and the limited data make detecting such power-laws a non-trivial task \cite{clauset2009power, marsili2013sampling, gerlach2019testing, serafino2021true}. Hence, it is paramount to understand whether the distribution of 2019-2020, characterized by a cutoff that is typically associated with the finite size of the system (Figure~\ref{fig:timeseries_properties}c), is quantitatively different from the ones of previous years. Notably, the range of fire sizes of previous years - e.g., 2002-2003 or 2006-2007 - show that fires larger than the ones in 2019-2020 took place, see Figure~\ref{fig:timeseries_properties}b and Figure~\ref{fig:timeseries_properties}d.

\begin{figure*}[t!]
    \centering
    \includegraphics[width = 0.8\textwidth]{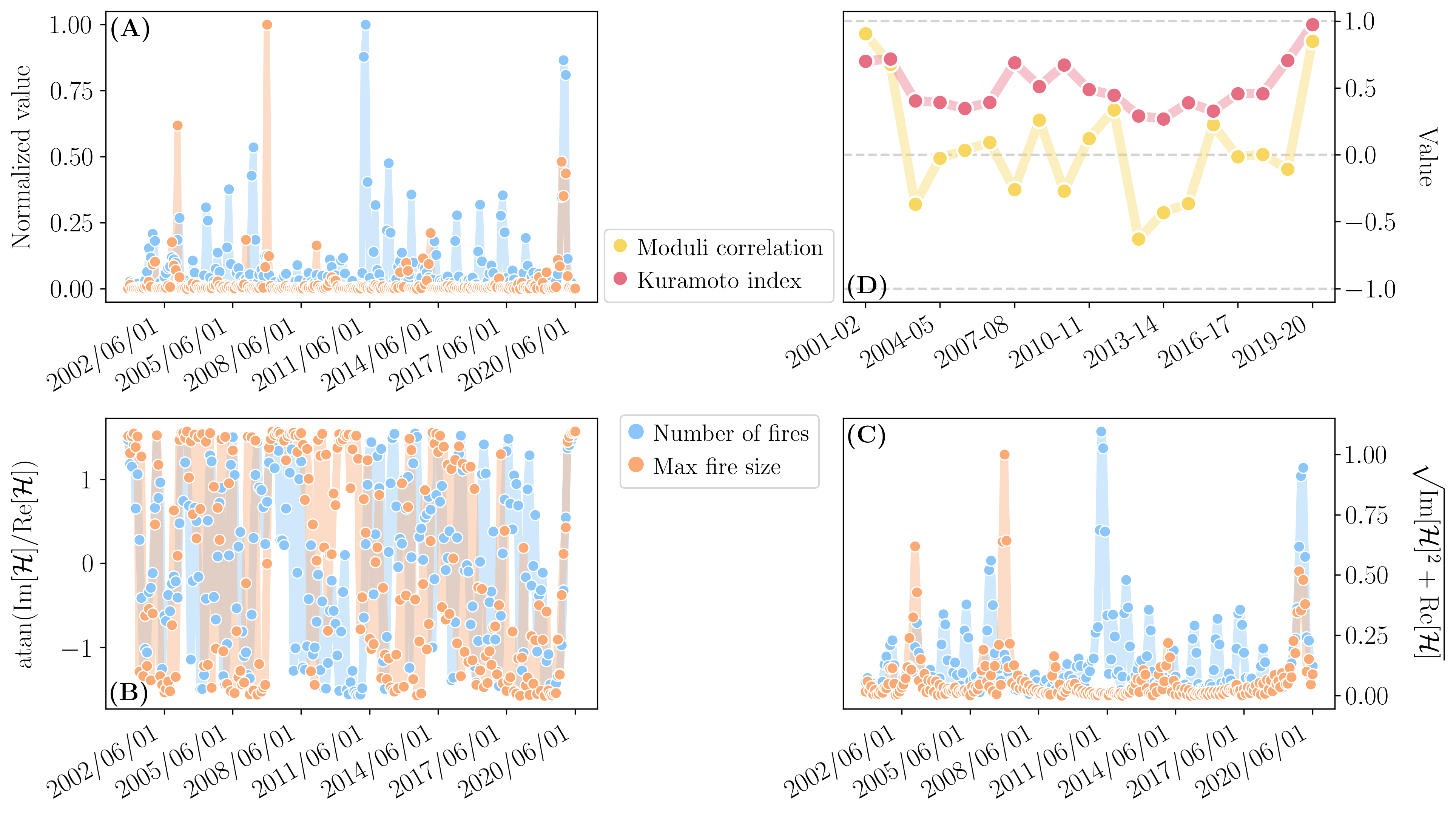}
\caption{\textbf{Properties of the timeseries of the number of fires and their maximum size.} (A) Plot of $N_{\rm fires}(t)$, the number of fire outbreaks in a given month, and of $M_{\rm fires}(t)$, the largest outbreak of a given month. Both timeseries have been normalized by their maximum value in order to compare them. (B) The phase of the Hilbert transform of both the timeseries of the number of fires and of the maximum fire size show a major synchronization during 2019-2020. (C) Similarly, the corresponding amplitudes suggest that during the 2019-2020 fire season a very large number of fires coexisted together with extremely large ones. This behavior is indeed captured by the power-law behavior of the fire size distribution. (D) The Kuramoto index between the phases of the Hilbert transform of $n_c(t)$ and of $m_c(t)$ and the correlation between the respective moduli. A clear synchronization emerges during 2019-2020, and in the same year the correlation between the moduli spikes as well. This behavior is compatible with the power-law cumulative distribution of the fire sizes found in the same year.}
\label{fig:max_num}
\end{figure*}

Therefore, to gain further insights into the fire dynamics, we extract the timeseries of the number of fires per month $N_{\rm fires}(t) = n_c(t)$ - i.e., the number of connected clusters of each matrix $M_t$ - and of the size of the largest fire $M_{\rm fires}(t)$ (i.e., the size of the largest connected cluster of each matrix $M_t$). In general, we do not expect these timeseries to be synchronized and, indeed, we typically see that a high number of clusters in a given year usually does not imply large clusters as well, as we can see in Figure~\ref{fig:max_num}a. To study the relation in time between $N_{\rm fires}(t)$ and $M_{\rm fires}(t)$, we perform a dynamical analysis of these two timeseries by introducing the phase and the modulus of their Hilbert transform (Methods). We plot them in Figure~\ref{fig:max_num}b-c. Then, for every year, we compute the associated Kuramoto index \cite{pikovsky2002synchronization}, which measures the synchronization between the number of fires and the maximum fire sizes, and the moduli correlation (see Figure~\ref{fig:max_num}d and Methods for further details). We find that during the 2019-2020 season Australia experienced neither the largest fire in our data nor the largest number of fire outbreaks - but rather a major synchronization between the two timeseries emerges. That is, both the number of fires and the size of the largest fires suddenly increased with respect to previous years. Such presence of many and very large outbreaks at once can be interpreted as a distinctive proxy of the widespread damage the outbreaks caused. Furthermore, it suggests that the distribution we see in 2019-2020 may be associated to features that we expect to see in power-law distributions and scale-free phenomena. Crucially, such a scale-free distribution might be tightly related to the dramatic impact that the 2019-2020 bushfire season had on the vegetation and on biodiversity. In fact, fire sizes that are distributed as a power-law give rise to both a few large fire clusters - corresponding to the distribution's long tails - and many smaller ones - the bulk of the power-law distribution. As the former devastated entire regions, the latter created pockets of vegetation fuel in other areas that could possibly act as an ignition to the next high-impact fire. Such a catastrophic departure from the historical trend suggests that a fundamental shift in the underlying dynamics might have occurred. In the next section, we will test the hypothesis that the 2019-2020 distribution is compatible with a scale-free one by introducing a spatial coarse-graining.

\subsection*{Spatial coarse-graining}
\noindent The spatial resolution of our data allows us to carefully test the hypothesis that the power-law distribution we see during 2019-2020 is a signature of an underlying scale-invariance. Here, we draw inspiration from the Renormalization Group concepts \cite{wilson1983renormalization, binney1992theory, loreto1995renormalization, goldenfeld2018lectures} and implement the so-called coarse-graining (CG) step to understand if such scale-invariance is present. We perform, at each time, a spatial CG through a block-spin transformation of $M_t$, by grouping together nearby pixels in $2 \times 2$ plaquettes. We then define the new super-pixels through a majority rule, in such a way that if the plaquette contained a majority of burned pixels, the corresponding coarse-grained pixel will be burned as well, and vice-versa (Methods).

Then, we follow the properties of the system along these repeated transformations. In fact, a coarse-graining transformation amounts to studying a system at different spatial scales. If the system is truly scale invariant, we expect that its properties will not change under repeated CG steps. Hence, and compatibly with the quality of the data, if the distribution of the fire size is a true power-law it will remain a power law after one or more CG transformations, with a corresponding finite-size scaling correction (Methods). In principle, one should iterate the coarse-graining indefinitely, to unravel the properties of its fixed points - however, with real data we are limited by the finite size of our system. Since each of the coarse-graining steps we are employing reduces the linear size of the system by half, after four CG steps we are left with a matrix that contains only $\approx 0.4 \%$ of the initial number of pixels. If only few but large fires are present in the original system, this coarse-grained version will be dominated by system-sized outbreaks. On the other hand, if many but small fires characterized the initial state, the coarse-graining transformations will drive the system to a configuration where virtually no fires are present. In particular, the behavior of probability distributions along the coarse-graining is particularly relevant in determining the properties of a critical system \cite{jona1975renormalization}.

\begin{figure*}[t!]
\centering
    \includegraphics[width = 0.85\textwidth]{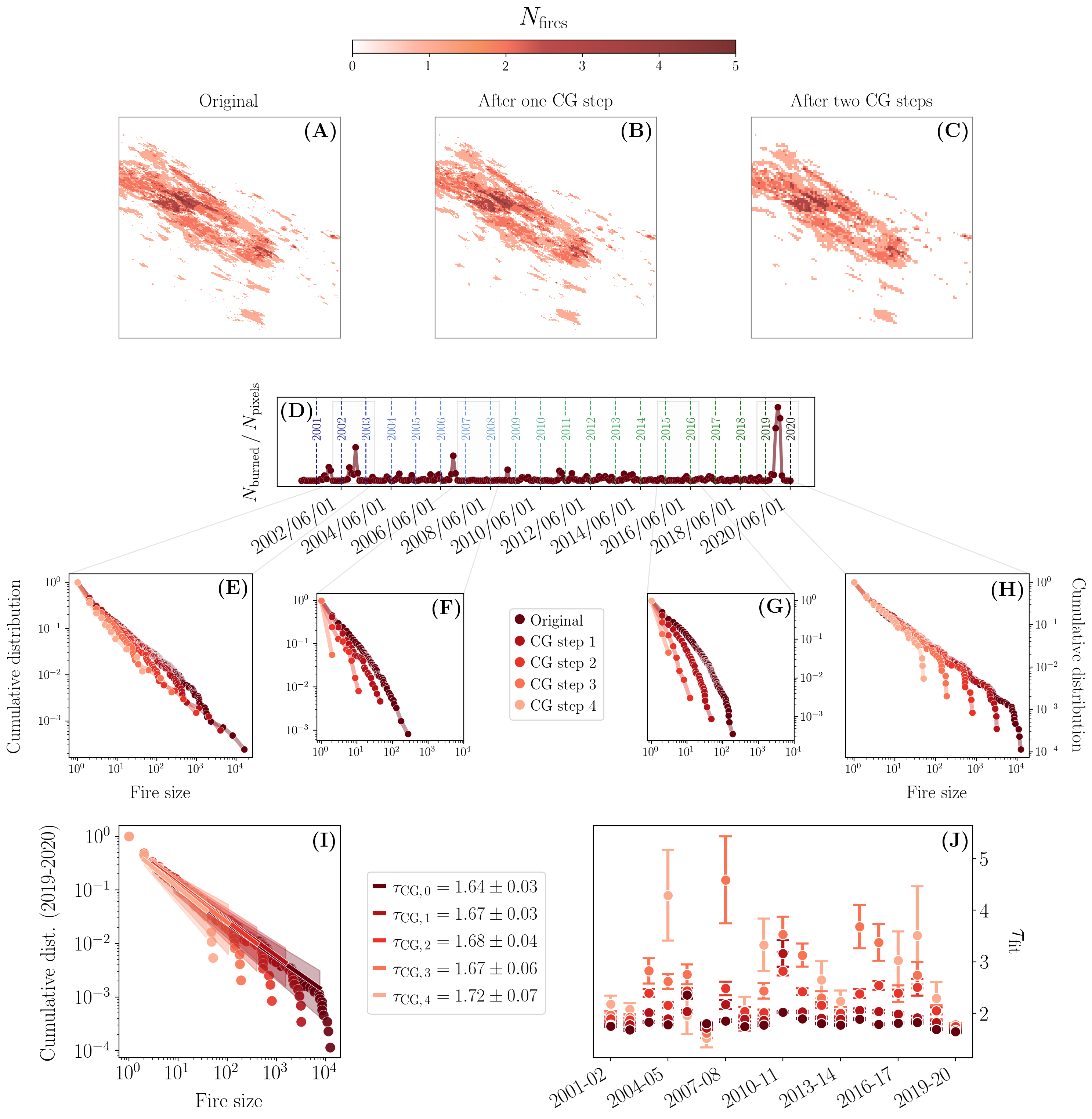}
\caption{\textbf{The properties of the data under spatial coarse-graining.} (A-C) An example of the effect of three coarse-graining steps on the overall number of fires per pixel in a sub-region of our data. The more steps are performed, the less the number of pixels left. (D) The timeseries of the burned area $N_{\rm burned}$ plotted in Figure~\ref{fig:timeseries_properties}, shown as a reference. (E-H) The coarse-graining corroborates the presence of a very robust scale invariance during 2019-2020, whereas in the previous years the shape of the distribution is significantly changed by the CG transformation. For instance, during 2007-2008 and during 2015-2016 the distribution of the fire sizes is exponentially suppressed, and after four coarse-graining steps there are almost no more fires to begin with. The distribution of 2002-2003, although not exponential, lacks the expected finite-size cutoffs and displays changes in the bulk of the distribution along the coarse-graining. (I) Power-law fit of the coarse-grained distributions in 2019-2020 using maximum-likelihood fitting methods. (J) If we fit a power-law to each year, we find that 2019-2020 displays the most consistent exponents at different coarse-graining steps. For comparison, in, e.g., 2006-2007 the exponents vary in the range $\approx[1.5,1.8]$, which is almost four times larger than the one found in 2019-2020. See also Figure S3.}
\label{fig:data_coarse_graining}
\end{figure*}

In Figure~\ref{fig:data_coarse_graining}a-c we show an example of the effect of three coarse-graining steps on a sub-region of the matrix $M = \sum_t M_t$ where each entry indicates the total number of times the corresponding pixel has been burned. As we can see, the spatial coarse-graining preserves some features of the original matrix, although the number of pixels is reduced by a factor $2^4$. As a reference, in Figure~\ref{fig:data_coarse_graining}d we show once more the timeseries of the burned area $N_{\rm burned}$ and in Figure~\ref{fig:data_coarse_graining}e-h we follow the cumulative distributions of the fire sizes along the coarse-graining for selected years.

For instance, during the 2007-2008 season or the 2015-2016 season the CG quickly suppresses the fire sizes distribution and only small fires are left. Interestingly, if we consider a season associated with a marked peak of the burned area, such as the 2002-2003 season in Figure~\ref{fig:data_coarse_graining}e, we see that, although the distribution keeps its distinctive long tails along the coarse-graining, the bulk of the distribution changes and no evident cut-off appears. On the other hand, and crucially, during the 2019-2020 season the bulk of the power-law distribution of the fire sizes is left invariant, while the cut-off associated with successive CG steps is poised at smaller and smaller system' sizes (see Methods).

Quantitatively, in Figure~\ref{fig:data_coarse_graining}i we show that maximum-likelihood fits \cite{clauset2009power} of the distributions at different levels of coarse-graining in 2019-2020 the exponents remain compatible with one another. Then, we repeat the same procedure for the other years and plot the corresponding exponents in Figure~\ref{fig:data_coarse_graining}j. Taking into account the standard deviation of these exponents, only the distribution associated with the 2019-2020 season consistently displays a power-law behavior with the same exponent at all CG steps.

These results strongly support the power-law nature of the 2019-2020 distribution, revealing a unique underlying scale-invariance in the spatial structure of the fire outbreaks taking place in that season. This scale-invariance, in turn, manifests itself dynamically as a synchronization between the number of outbreaks and their maximum size. Altogether, our data suggest that the emergent properties we observe are related to a phase transition. Thus, a fundamental question arises: what has driven the 2019-2020 fire dynamics close to what appears to be a critical point?

\subsection*{Paradigmatic model for the vegetation-fires dynamics}
\noindent To qualitatively understand the abrupt changes observed during the 2019-2020 bushfire season, we introduce a minimal stochastic model of the forest-fire class \cite{drossel1992self}. Differently from classical forest-fire models that display self-organized criticality \cite{bak1990forest, malamud1998forest, bak2013nature} and multistability \cite{PhysRevE.104.L012201}, we describe the concurrent stochastic spread of both fires and vegetation between neighboring nodes of a network. Although extensions of the forest-fire model with different vegetation growth and with climate effects have been proposed \cite{pueyo2007self, staal2018resilience}, our approach seeks to include only minimal features to understand whether they are able to explain the patterns observed in our data. Without fires, the vegetation $V$ is free to spread to its nearest neighbors at a rate $\lambda_V$ on a given graph - for instance, a 2-dimensional lattice - and spontaneously disappears with a death rate $d_V$. Then, a fire $F$ can ignite on a vegetation site with rate $b_F$ and spreads with a rate $\lambda_F$ over an effective topology that is determined by the structure of the vegetation clusters. At the same time, the vegetation cannot occupy a site with a fire $F$, thus both the topology of the fire layer and of the vegetation layer change dynamically with time. Once a fire is over, with a rate $d_F$, the corresponding site will become an empty site $\varnothing$ for the vegetation layer, and will not be present in the fire layer. Hence, although archetypal, our model is described in terms of few parameters that can be thought of as functions of environmental conditions.

\begin{figure*}[t!]
\centering
    \includegraphics[width = 0.9\textwidth]{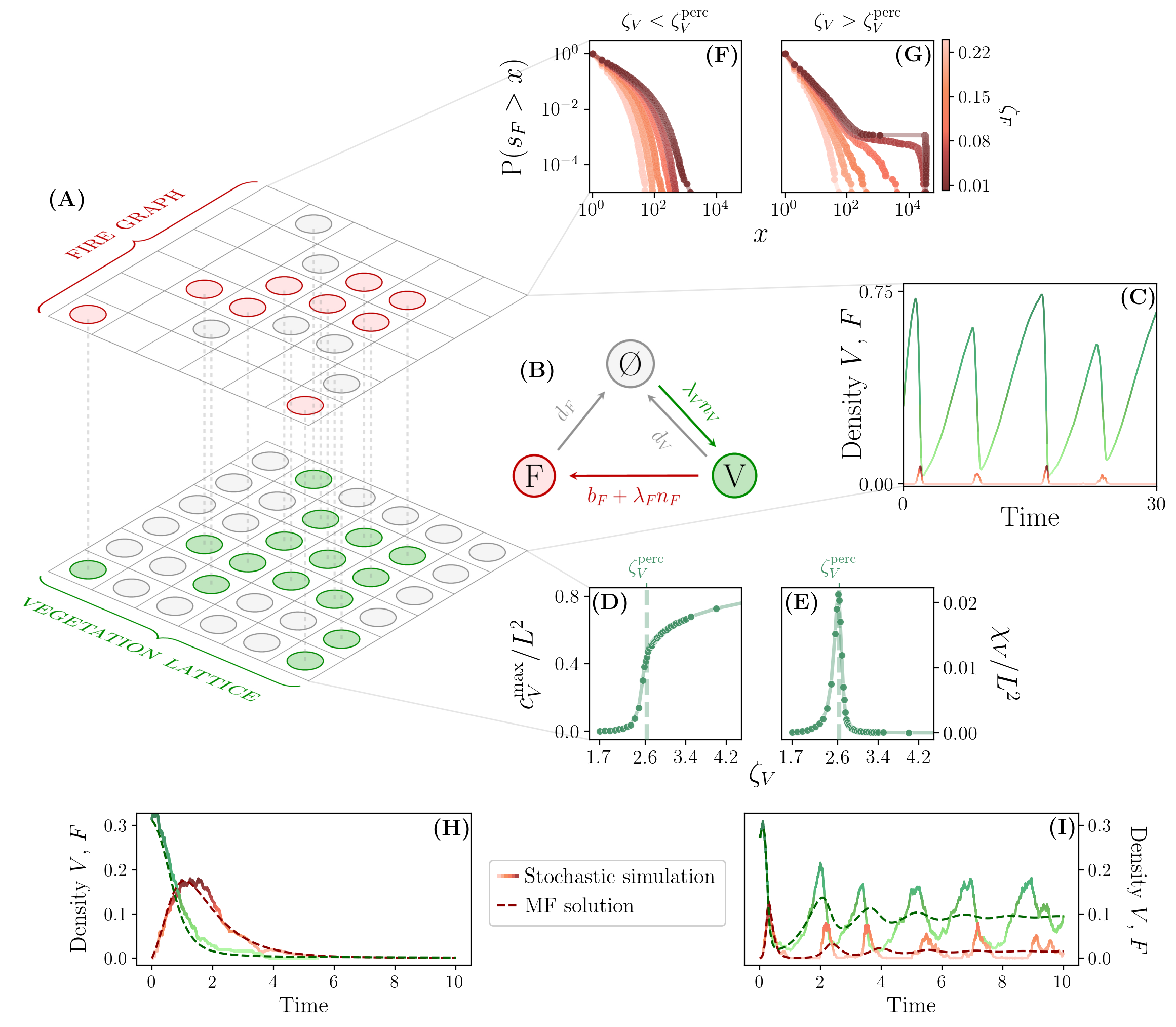}
\caption{\textbf{The fundamental properties of our model.} (A-B) A depiction of the model dynamics as a multi-layer graph and the corresponding transition rates. (C) On a $2D$ lattice the model displays a charge-discharge behavior if the vegetation dynamics is much slower than the fire one, and fires are relatively rare events. Here $(d_F, b_F, \lambda_F) = (25, 10^{-5}, 500)$ and $(d_V, \lambda_V) = (0.5, 3)$. (D-E) The vegetation layer undergoes an isotropic percolation transition at $\zeta_V^{\rm perc} \approx 2.63$ where a spanning cluster appears. In (D) we plot the size of the largest vegetation cluster $c_V^{\rm max}$ and in (E) the mean vegetation cluster size $\chi_V$, which peaks at the transition. Both the plots are from a $250\times250$ lattice (see Figure S4). (F-G) If we consider fires that spread over a fixed vegetation configuration (Methods), (F) below the percolation threshold $\zeta_V<\zeta_V^{\rm perc}$ the cumulative distribution of the fire size $s_F$ is always exponentially suppressed due to the small vegetation clusters. Above it (G), the fires may spread on a spanning cluster, and therefore we have system-size outbreaks if $\zeta_F$ is small enough. (H-I) Comparison of the analytic solution of the mean field equations and a stochastic simulation on a fully connected network with $500$ nodes. (H) With parameters $(d_F, b_F, \lambda_F) = (1, 0.5, 10)$ and $(d_V, \lambda_V) = (0.1, 0.5)$ the absorbing state, i.e., the empty configuration is the stable mean-field solution. (I) For $(d_F, b_F, \lambda_F) = (10, 0.1, 100)$ and $(d_V, \lambda_V) = (1, 3)$, instead, noise-induced oscillations around the mean-field stationary values emerge. Notably, the mean-field approximation is not able to predict the charge-discharge behavior described above, which is a consequence of spatial effects that are thus fundamental in our model.}
\label{fig:model}
\end{figure*}

Notably, a similar model was proposed by Zinck et al. \cite{zinck2011shifts} to analyze data from the Canadian Boreal Plains. However, the modeling approach proposed by the authors did not include nearest-neighbor spreading nor a death rate for the vegetation. As we will see, in our model, these parameters are fundamental in shaping the spatial structure of fires. Indeed, heuristically, we can think of our model as defined on a multi-layer network \cite{de2015structural, de2016physics, kivela2014multilayer}. In this depiction, the topology of the vegetation layer is fixed, but the vegetation sites dynamically govern the topology of the fire layer, as we sketch in Figure~\ref{fig:model}a. Hence, we expect the interplay between the spatial spreading of both vegetation and fires to be a crucial feature of our model, whose rates are shown in Figure~\ref{fig:model}b. The vegetation alone obeys
\begin{equation}
\label{eq:reactions_vegetation}
\begin{gathered}
    V_i + \varnothing_{j \in \partial i} \xrightarrow{\lambda_V} V_i + V_j \\
    V_i \xrightarrow{d_V} \varnothing_i,
\end{gathered}
\end{equation}
where $i$ is a site and $\partial i$ is the set of the neighbors of $i$. These reactions for the vegetation dynamics, thus correspond to the well-known contact process \cite{harris1974,dickman1999,marro2005nonequilibrium,henkel2009}, an archetypal model of absorbing phase transitions. We highlight that, differently from most SOC and previous models, we do not include an immigration term for the vegetation - i.e., an external field in the contact process. This amounts to assuming that vegetation can only spread from other vegetation sites, rather than reappearing in random sites. On top of this dynamics the fire spreading is determined by the reactions
\begin{equation}
\begin{gathered}
    F_i + V_{j \in \partial i} \xrightarrow{\lambda_F} F_i + F_j \\
    V_i \xrightarrow{b_F} F_i \\
    F_i \xrightarrow{d_F} \varnothing_i.
\end{gathered}
\end{equation}
These reactions, if considered independently from Eq.~\eqref{eq:reactions_vegetation}, represent instead a contact process with resource depletion - meaning that the empty sites are unavailable for fires to spread.

\subsection*{Model simulations and time-scale separation}
\noindent We perform exact stochastic simulations of the model on a $2$-dimensional lattice using the Gillespie algorithm (Methods) \cite{gillespie1977exact}. Crucially, for the model to be reasonable, we must assume that the vegetation dynamics is much slower than the one of the fires and that the birth rate of the fires $b_F$ is typically very small \cite{zinck2011shifts}. In Figure~\ref{fig:model}c we show that the model in this range of parameters indeed displays a charge-discharge dynamics, with long periods of almost undisturbed vegetation spreading followed by shorter periods of fire spreading following the rare ignition of an outbreak.

This time-scale separation limit corresponds to the assumption that the vegetation configuration does not change during the propagation of a fire. Therefore, we study how a fire propagates on top of a fixed stationary vegetation configuration. In this scenario, the phase space is described by the adimensional parameters $\zeta_F = d_F/\lambda_F$ and $\zeta_V = \lambda_V/d_V$ (Methods). A small value of $\zeta_F$ gives rise to fires that are extremely effective at spreading and, vice-versa, a large value of $\zeta_V$ implies a quick vegetation regrowth. Remarkably, since the vegetation layer in the absence of fires follows a simple contact process, we expect a percolation transition at $\zeta_V^\mathrm{perc} \approx 2.6$, as recently shown with numerical simulations \cite{martin2020intermittent}. At this value, a system-size cluster of vegetation appears, coexisting with a significant number of distinct, but smaller vegetation clusters (Figure~\ref{fig:model}d-e). At $\zeta_V = \zeta_V^{\rm perc}$, in particular, if $L\to\infty$ an infinite cluster of vegetation appears. If we call $c_V$ the vegetation cluster size and $n(c_V)$ the number of vegetation clusters of size $c_V$, we can define the mean vegetation cluster size ratio between the first two moments as
\begin{equation}
    \chi_V =  \frac{\sum_{c_V} (c_V)^2 \, n(c_V)}{\sum_{c_V} c_V \, n(c_V)}
\end{equation}
where the sum runs over all vegetation clusters. This quantity, which we plot in Figure~\ref{fig:model}e, is expected to diverge at the percolation transition due to the scale-free nature of the vegetation cluster sizes. In fact, in percolation theory, this is nothing but the mean cluster size $\chi = \sum_{c_V} c_V w_{c_V}$, with $w_{c_V} = c_V n(c_V) / \sum_{c_V} c_V n(c_V)$ the probability that a vegetation site belongs to a cluster of size $c_V$, which displays a power-law divergence close to the percolation threshold \cite{stauffer2018introduction}.

This transition has a crucial impact on the cumulative distribution of the fire sizes $s_F$, as we see in Figure~\ref{fig:model}f-g. In fact, below the percolation transition of the vegetation, fires are severely limited by the size of the vegetation clusters, and thus the distribution of $s_F$ is exponentially suppressed even at small $\zeta_F$. On the other hand, above the percolation threshold, the vegetation clusters tend to be larger, and fires can be large if $\zeta_F$ is small enough. This suggests, as highlighted in \cite{zinck2011shifts}, that a critical transition may underlie the vegetation-fires dynamics.

Although this percolation-like transition emerged from the spatial nature of our model, we can also solve it analytically in a mean-field approximation - which amounts to ignoring such spatial features to begin with (Methods). Yet, the mean-field solution allows us to reveal the presence of yet another critical point, an absorbing phase transition \cite{dickman1999,marro2005nonequilibrium,henkel2009}. This phase transition separates a phase in which the mean-field stable configuration predicts a non-zero density of both fire and vegetation from a phase in which the stable configuration is the empty one, see Figure~\ref{fig:model}h-i. Crucially, the mean-field picture is drastically different from a spatially embedded model. Indeed, the spatial structure significantly changes the way fires spread due to the modulation of the underlying vegetation structure, leading to isotropic percolation which will play a fundamental role.

\subsection*{Model coarse-graining and the emergence of scale-free fires}
\begin{figure*}[t!]
\centering
    \includegraphics[width = 0.95\textwidth]{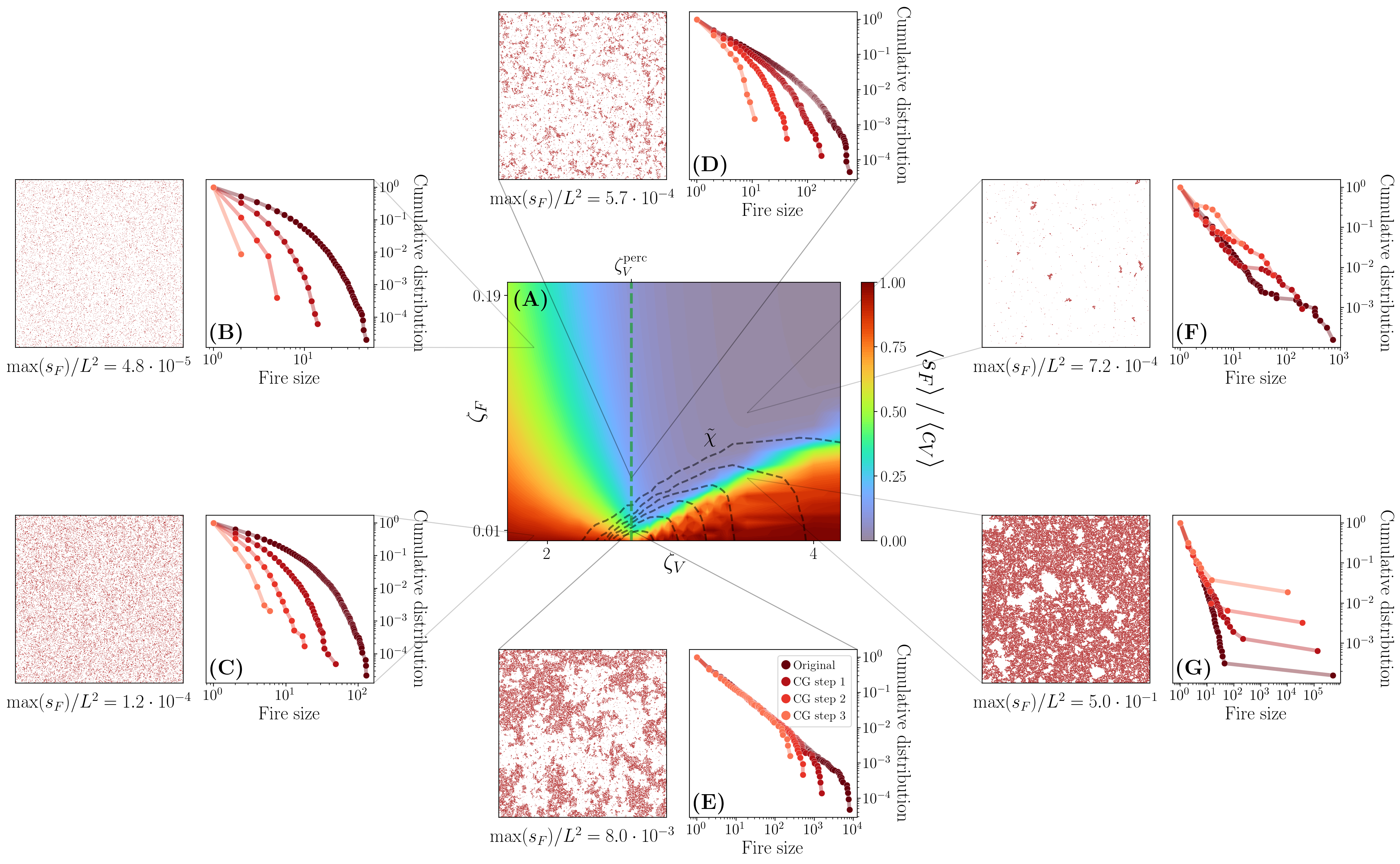}
\caption{\textbf{The properties of the time-scale separated model and its behavior under spatial coarse-graining.} (A) At a given set of parameters $(\zeta_V, \zeta_F)$ we plot the ratio between the mean fire size $\ev{s_F}$ and the mean size of a vegetation cluster $\ev{c_V}$ in a $250\times 250$ lattice. The black dotted lines represent contour lines of $\tilde{\chi}$ (the product of the number of vegetation clusters $n_{c_V}$ and the maximum fire size $s_F^{\rm max}$), which is maximized around the percolation transition $\zeta_V^\mathrm{perc}$ for low enough values of $\zeta_F$. (B-G) We seed $n_F^0 = 10^5$ fires on a lattice with linear size $L = 1000$ in order to study the distribution of the fire sizes $s_F$ and the corresponding coarse-grained distributions. (B-C) At low values of $\zeta_V$, the cumulative distribution of the fire sizes is exponential and is further suppressed along the coarse-graining at all values of $\zeta_F$. (D-E) At $\zeta_V^\text{perc}$, if $\zeta_F$ is low enough the fire size distribution becomes a power-law that is invariant along the coarse-graining. (F-G) For high values of $\zeta_V$, on the other hand, the system is dominated by few large clusters of vegetation, and the corresponding large fires are highlighted by the coarse-graining. This regime is not particularly realistic at low $\zeta_F$, since it would require climate conditions that allow for large fires, i.e., a warm and arid climate, but at the same time for an extremely effective vegetation spread.}
\label{fig:order_parameter_distributions}
\end{figure*}

\noindent Simulations of the model allow us to study the properties of the area burned by fires at different values of $(\zeta_V, \zeta_F)$. In particular, in Figure~\ref{fig:order_parameter_distributions}a we show the behavior of the ratio between the average fire size $\ev{s_F}$ and the average vegetation cluster size $\ev{c_V}$ observed in configurations with given parameters $(\zeta_V, \zeta_F)$. This parameter is fundamental because it helps us understand the potentially damaging effects of the fires on the underlying vegetation substrate. Whenever $\ev{s_F}/\ev{c_V} \approx 1$, it implies that a fire that originates in a given vegetation cluster has a non-vanishing probability to burn the entire cluster.

In Figure~\ref{fig:order_parameter_distributions}a we plot another relevant quantity as well - the black dotted lines represent the contour lines of $\tilde{\chi} = s_F^{\rm max} \times n_{c_V}$, where $n_{c_V}$ is the number of vegetation clusters. This quantity is particularly significant because $n_{c_V}$ can be interpreted as a rough estimate of the number of possible fires in the system, whereas $s_F^{\rm max}$ tells us how large they can be. In the data, these two quantities both reached high values at the same time during 2019-2020.

In order to study the behavior of the fire sizes distribution under the same spatial coarse-graining applied in the data, we choose $n^0_F = 10^5$ fire seeds in a large lattice of linear size $L = 1000$ (Methods). Then, we analyze the resulting burned area in a given point of the phase space $(\zeta_V, \zeta_F)$. In particular, we look at the distribution of the fire sizes and, thanks to the large size of the lattice, at how it changes along repeated CG transformations. We find four different regimes, shown in Figure~\ref{fig:order_parameter_distributions}a-g. If $\zeta_V$ is high enough, typically the vegetation can spread effectively and regrow any burned vegetation. Yet, if $\zeta_F$ is low, fires can propagate almost unboundedly due to the underlying large vegetation clusters. The resulting distributions, shown in Figure~\ref{fig:order_parameter_distributions}g), are therefore dominated by very large fires. Indeed, since for $\zeta_V > \zeta_V^{\rm perc}$ a spanning cluster is present, vegetation sites that are far away are likely connected and fire can spread from one to the other. This regime is perhaps unrealistic since it leads to extremely large fires in an otherwise vegetation-rich environment. Yet, a similar dynamics was observed in fire-prone communities where species with post-fire recruitment have the most flammable canopies \cite{Keeley2011}.

On the other hand, a more realistic regime is described by a high $\zeta_V$ and a high $\zeta_F$ as well. This regime corresponds to environmental conditions that favor a vegetation-rich system and suppress fires, and therefore we expect to see a small burned area. In fact, as we see in Figure~\ref{fig:order_parameter_distributions}f, fires are small as they are not able to propagate effectively, not even on the underlying spanning cluster of vegetation sites. Crucially, in both these regimes  (Figure~\ref{fig:order_parameter_distributions}f-g) the coarse-graining accentuates the tails of the fire size distribution, since the coarse-graining will unravel the largest fires that propagate on the vegetation spanning cluster.

On the other hand, if $\zeta_V$ is low, vegetation regrowth is typically suppressed. In this case, when $\zeta_F$ is high, fires tend to be small as we see in Figure~\ref{fig:order_parameter_distributions}b, but so do the clusters of vegetation. Indeed, $\ev{s_F}/\ev{c_V}$ can dangerously increase because substantial parts of the underlying vegetation clusters can burn even at high $\zeta_F$. Finally, when $\zeta_F$ is also low, not only the vegetation clusters can hardly regrow, but a fire can systematically burn the entire cluster in which it originates since $\ev{s_F}/\ev{c_V} \approx 1$. This regime is not sustainable in the long time - the fires are likely to outpace the vegetation regrowth and eventually desertification will take place. Notably, this regime cannot be distinguished from the distribution of the fire sizes alone, in Figure~\ref{fig:order_parameter_distributions}c, which stays exponential due to the lack of large vegetation clusters.

The vegetation percolation transition lies in between these regimes, and it is here that power-law distributed fires emerge at low enough $\zeta_F$ (Figure~\ref{fig:order_parameter_distributions}d-e). In fact, at this point we see a distinctive scale-invariant configuration emerging following a power-law distribution with an invariant bulk under spatial coarse-graining. Moreover, this is also the region where $\max{s_F}\times n_{c_V}$ is maximized, since the system can experience large fires coexisting with a large number of clusters of vegetation. Therefore, the features that we observe during the 2019-2020 fire season are best described by our model in a time-scale separation approximation close to the vegetation percolation transition $\zeta_V = \zeta_V^{\rm perc}$, with small fire suppression $\zeta_F$. Note that in this regime, $\ev{s_F}/\ev{c_V}$ remains small. However, the mean itself is not representative in the presence of scale-free distributions, hence it is not a reliable index of fire damage anymore.

\section*{Discussion}
\noindent How did Australia reach such a critical point? Although our modeling approach is paradigmatic, it provides a clear physical interpretation of its control parameters $\zeta_V$ and $\zeta_F$. Indeed, their value is determined by climate conditions. Therefore, prolonged droughts, higher temperatures, and a more arid climate - all recognized as contributors to the 2019-2020 bushfire season \cite{dowdy2018climatological, arriagada2020climate, abram2021connections} - might have pushed both $\zeta_V$ and $\zeta_F$ to lower and lower values, eventually reaching and crossing the percolation transition between 2019 and 2020. Notably, the 2019-2020 year has been unusually hot and dry in part due to natural meteorological phenomena, such as a shift in the polar winds above Antarctica and one of the strongest positive swings in the Indian Ocean Dipole. The former contributed to stratospheric warming, which in turn contributed to bringing hot, dry weather to much of Australia. The latter, in its positive phase, may have led to a reduction in rainfall over the southern and most northerly regions of Australia \cite{dowdy2018climatological}. However, on top of - and possibly as a cause of - all this natural variation, global warming is making the country even hotter and drier \cite{jyoteeshkumar2021intensifying}, with the devastating effects that we highlighted in this work.

Finally, the mean-field analysis and the vegetation layer of our paradigmatic model predict the presence of yet another critical point, one of a very different nature associated with the absorbing phase transition of the contact process \cite{harris1974,dickman1999,marro2005nonequilibrium,henkel2009} at $\lambda_V/d_V := \zeta_V^\text{abs} \approx 1.6$. This phase transition separates a phase in which the only stable configuration is the absence of vegetation, and a phase in which vegetation is present \cite{Fairman2015}. Crucially, with the addition of the fire dynamics, a slow enough vegetation spreading implies that fires at high values of $\zeta_F$ can burn large clusters of vegetation. This scenario may push the system to a state in which the vegetation goes extinct. Such states are much harder to reach in more realistic and highly complex dynamics of fire spreading in forests. For example, one should consider that broadleaf Australian forest species, such as Eucalyptus, have resilience and resistance traits, like re-sprouting and seed banks, that allow for a rapid post-fire recovering even in intense fire-regimes \cite{steel2021ecological, nolan2021limits}. Yet, repeated fires with short return times would cause the exhaustion of these capacities \cite{Fairman2015}. These considerations do suggest that the isotropic percolation transition observed during the 2019-2020 bushfire crisis may foreshadow a worsening condition that, in the far future, might push the system to a forest-savanna-like type of transition \cite{de2013fire,oliveras2016many}.

Overall, our results suggest that the unprecedented bushfire season that Australia experienced between 2019 and 2020, with outbreaks appearing at all scales, is compatible with a phase transition in the vegetation-fires dynamics driven by a worsening climate. Our work shows how phase transitions and critical points play a fundamental role in shaping this dynamics, and their presence and consequences will be more and more relevant as climate change will quickly deteriorate the climatic conditions.

\section*{Limitations of the study}
\noindent Future works should aim to develop quantitative methods to infer the values of the model's parameters from data, both from both fires spreading and vegetation evolution. Although the present study lacks such inference steps, procedures such as simulation-based inference \cite{cranmer2020frontier} may be well-suited to this aim. In particular, ecological and environmental drivers evolve over time, both due to seasonality and climate change. This would amount to prescribe a dynamics for the parameters $\zeta_F$ and $\zeta_V$ of our model, as well as $b_F$, which are instead considered constant in our analysis. Changes in these parameters over time may affect the dynamics, creating feedback effects that are taken into account in the present study. Notably, quantities from Information Theory may be a promising extension to disentangle the environmental effects from the vegetation-fire interaction \cite{nicoletti2021mutual}. It will be crucial to account for and disentangle contributions coming from natural variations and from the anthropogenic impact, in order to assess mitigation strategies that are becoming more and more vital. Furthermore, here we only apply from the Renormalization Group in the form of coarse-graining and finite-size scaling. It will be of particular interest to consider other phenomenological approaches to the Renormalization Group, beyond simple coarse-graining \cite{nicoletti2020scaling}. Finally, it will be paramount to apply the analysis carried out in this work to other areas of the world where large and extended fire outbreaks are appearing.

\section*{Acknowledgments}
\noindent AM was supported by ``Excellence Project 2018'' of the Cariparo foundation. LS was financially supported by "Fondo para la Investigación Científica y Tecnológica (FONCYT) PICT 2020 SERIE A 2628".

\section*{Author contributions}
\noindent L.S. and S.S. designed the study. L.S., A.M., and S.S. supervised the research. L.S. and F.M. collected the data. G.N. analyzed the data. G.N., A.M., and S.S. designed and studied the model. G.N. performed the simulations and the analytical calculations. G.N. prepared the figures. G.N. and S.S. interpreted the results and wrote the first draft of the manuscript. All authors contributed to the final version of the article and approved the submitted version.

\section*{Declaration of interests}
\noindent The authors declare no competing interests.

\section*{Data and code availability}
\noindent This paper analyzes existing, publicly available data from the NASA MODIS platform. All original code has been deposited at Zenodo and is publicly available at \cite{giorgio_nicoletti_2023_7540390, lsaravia_2023_7541674}.

\appendix
\clearpage
\newpage
\section*{METHODS}
\subsection*{Data collection}
\noindent We defined the region of study as the East and Southeast temperate broadleaf and mixed forests of continental Australia using the ecoregions defined by Dinerstein \cite{Dinerstein2017}, accessible at http://ecoregions2017.appspot.com/, which represents an area of $48 \cdot 10^6$ha (see Figure S1). For this region, we estimate the burned areas using the NASA Moderate-Resolution Imaging Spectroradiometer (MODIS) burnt area Collection 6 product MCD64A1 \cite{Giglio2016}, which is a monthly product with a $500$m pixel resolution. We downloaded the images, using Google Earth Engine, as geoTIFF and then we converted them to a binary matrix (circa $4000x8000$) using the R statistical language \cite{RStat2021}. Then, for each month we have a binary matrix $M_t$, whose pixels represent an area of $500\, \text{m}^2$ and can be either $1$ - if there has been a fire in that pixel in the span of that month - or $0$ - if no event occurred, meaning that no burned area was detected.

\subsection*{Cluster distributions}
\noindent We define a cluster of a binary matrix $M_t$ using a nearest-neighbors connectivity, i.e. the pixels that belong to a cluster are defined using the connectivity matrix
\begin{equation}
    C_\text{basis} = \begin{pmatrix}
                        0 & 1 & 0 \\
                        1 & 1 & 1 \\
                        0 & 1 & 0 \\
                     \end{pmatrix}
\end{equation}
which defines the usual nearest neighbors of a $2$-dimensional lattice. We also repeated the analysis described in the main text using a next-nearest-neighbors connectivity and the results do not change significantly. Therefore, for each matrix $M_t$ we end up with a number of clusters $n_c(t)$ and the areas of each cluster $\{A_c^{(i)}\}_{i=1}^{n_c(t)}$. Then the cumulative fire size distribution of $M_t$ can simply be evaluated as
\begin{equation}
    P(s_F) := P(A_c > s_F) = \sum_{i = 1}^{n_c(t)}\frac{\theta\left(A_c^{(i)} - s_F\right)}{n_c(t)}
\end{equation}
where $\theta(\cdot)$ is the Heaviside function. To evaluate yearly distributions, we pooled the cluster sizes from all matrices $M_t$ of a given year. This amounts to assuming that the clusters found in subsequent months are independent. Indeed, we find that the overlap between burned pixels in $M_t$ and $M_{t+1}$ is always small, with respect to the number of burned pixels.

\subsection*{Cluster dynamics}
\noindent We can exploit the timeseries of both the number of clusters and their areas to probe the underlying properties of the fire dynamics. In particular, we look at the number of clusters $n_c(t) = N_{\rm fires}(t)$ and the area of the largest cluster $m_c(t) = \max_i \{A_c^{(i)}\}_{i=1}^{n_c(t)} = M_{\rm fires}(t)$. We normalize both these timeseries by dividing them by their maximum value, in order to make them comparable (Figure~\ref{fig:max_num}a). Other normalizations, such as a standard z-score, give essentially the same results. In order to understand how the evolution of these two timeseries relates in time, we introduce the Hilbert transform of a real-valued timeseries $x(t)$ as
\begin{equation}
    \mathcal{H}[x(t)] = x(t) + \frac{i}{\pi}\lim_{\varepsilon \to 0}\int_\varepsilon^\infty \frac{x(t+\tau)-x(t-\tau)}{\tau}d\tau
\end{equation}
which is a complex timeseries. Thus we can compute its phase $\varphi_x(t) = \arctan \frac{\Im[\mathcal{H}[x(t)]]}{\Re[\mathcal{H}[x(t)]]}$ and its modulus $\rho_x(t) = \sqrt{\Im^2[\mathcal{H}[x(t)]] + \Re^2[\mathcal{H}[x(t)]]}$ and how they change in time (Figure~\ref{fig:max_num}b-c).

We can further quantify the relations between $n_c(t)$ and $m_c(t)$ by looking at the Kuramoto index \cite{pikovsky2002synchronization} of their Hilbert transforms and at the correlation between the corresponding moduli. We define the Kuramoto index on a given year as
\begin{equation}
    K_\text{year} = \bigg|\ev{e^{\varphi_{n_c}(t) - \varphi_{m_c}(t)}}_\text{year}\bigg|
\end{equation}
and the correlation between the moduli as
\begin{equation}
    C_\text{year} = \frac{\ev{\rho_{n_c}\rho_{m_c}}_\text{year}-\ev{\rho_{n_c}}_\text{year} \ev{\rho_{m_c}}_\text{year}}{\sqrt{\prod_{i\in\{m_c, n_c\}}\left[\ev{\rho_{i}^2}_\text{year}-\ev{\rho_{i}}^2_\text{year}\right]}}.
\end{equation}
In Figure~\ref{fig:max_num}d we do indeed see that $K_\text{year}$ becomes significantly close to $1$ during 2019-2020, hence the two timeseries are highly synchronized during the year. Similarly, the correlation between the moduli has a positive spike in the same period. It is worth noting that this is true even if in the original timeseries neither the number of clusters nor the size of the largest one are maximal during 2019-2020. The fundamental change in the behavior of the system is that, during this year, both of them peak in a synchronized fashion, which leads to the power-law distribution shown in the main text.

\subsection*{Spatial coarse-graining}
\noindent A quantitative and powerful way to assess the scale-invariance of a system is given by a properly defined coarse-graining procedure \cite{wilson1983renormalization, binney1992theory, goldenfeld2018lectures}. In the spirit of Statistical Physics, a suitable coarse-graining for a binary matrix $M_t$ is a block-spin transformation of the associated $2$-dimensional square lattice. Namely, the $k$-th coarse-graining step amounts to define a super-pixel $\sigma_{i'}^{(k+1)}$ from the previous pixels $\sigma_i^{(k)}$ via the majority rule
\begin{equation}
    \sigma_{i'}^{(k+1)} = \begin{cases} 
                1 & \text{if} \quad \sum_{j \in B_i} \sigma_i^{(k)} > \lfloor \text{card} (B_{i})/2 \rfloor\\
                0 & \text{otherwise}
               \end{cases}
\end{equation}
where $\lfloor \cdot \rfloor$ is the floor function and $B_i$ is the $i$-th set of pixels such that $\{B_i\}$ forms a non-overlapping covering the original $2$-dimensional lattice. In particular, we take $B_i \in \mathbb{M}(2\times 2)$ so that at each coarse-graining step the number of pixels is reduced to a fourth of the original ones and therefore we can perform enough coarse-graining steps. Notice that, in this case, the majority rule is not exact since the cardinality of $B_i$ is even. Thus, if $ \sum_{j \in B_i} s_i^{(k)} = 2$ we randomly assign the value of $s_{i'}^{(k+1)}$ to be either $0$ or $1$.

In the spirit of the Renormalization Group, we should follow physical observables and - in particular - probability distributions \cite{jona1975renormalization} to look for scale-invariance along the coarse-graining. That is, if the system is scale-invariant in a spatial sense we should see that, even if we are coarse-graining the system, some of its properties will not change up to some finite-size cutoff, because the small-scale features are indistinguishable from the large-scale ones. This is exactly what we look for when we compare the cumulative probability distributions of the cluster sizes at different coarse-graining steps. 

As at each coarse-graining step we observe a smaller and smaller system, we can exploit finite-size scaling. Thinking of a percolation-like transition \cite{stauffer2018introduction}, the probability distribution of the fire sizes in a system of linear size $L$ scales as
\begin{equation}
    P_{\rm cumulative}(s_F) = s_F^{-\tau+1} \psi\left(\frac{s_F}{L^D}\right)
\end{equation}
where $D$ is related to the critical exponent of the correlation length and $\tau$ is the exponent of the power-law distributed fire sizes. In particular, $D$ is the fractal dimension of the fires. Hence, for a properly chosen value of $D$, we expect that $P_{\rm cumulative}(s_F) s_F^{\tau-1}$ as a function of $s_F/L^D$ will collapse onto the same curve. We find this collapse with $D \approx 1.95$, which suggests once more, and in terms of the Renormalization Group, that the 2019-2020 fire seasons appear to behave like a system close to a phase transition (see Figure S3). In fact, the fractal dimension tells us the size $s_F$ of a fire outbreak changes with its linear size, i.e., $s_F \sim L^D$.

Notice that in bond percolation we would expect the fractal dimension to be $D = 91/48 \approx 1.896$  in a two-dimensional lattice, which is compatible with what we find in the data. However, the exponent $\tau$ of the fire size distribution is different from the one expected in bond percolation, suggesting that the universality class might be different. Let us note that, in our model, bond percolation only happens in the isolated vegetation layer, and not in the layer where fire propagates.

\subsection*{Contact process and critical points}
\noindent The contact process \cite{harris1974,dickman1999,marro2005nonequilibrium,henkel2009} is an archetypal model for absorbing phase transitions, which describes spreading phenomena over a set of sites $\{\sigma_i\}_{i = 1, \dots, N}$. Each site can be either occupied $\sigma_i = 1$ or empty $\sigma_i = 0$. Empty sites are occupied by neighboring occupied sites at a rate $\lambda$, whereas occupied sites become empty at a rate $\mu$. The mean-field equations for the density of occupied sites $\rho$ is given by $\dot{\rho} = \rho(\lambda-\mu) - \lambda\rho^2$.

This equation has two stationary solutions. The first one is the empty configuration $\rho_{\rm st}^{\rm v} = 0$, which is only stable if $\lambda < \mu$. The empty configuration is an absorbing configuration, that is, once it is reached the system cannot leave it since no reactions are possible. If $\lambda > \mu$, the stable stationary solution is $\rho_{\rm st}^{\rm a} = 1 - \mu/\lambda$. The value $\lambda_{\rm abs} := \mu$ is the critical point of the model at which the absorbing phase transition takes place - below $\lambda_{\rm abs}$, the system will always reach the absorbing empty configuration, whereas above $\lambda_{\rm abs}$ non-zero values of $\rho$ are possible. Conversely, $\rho$ is the order parameter of the system, which identifies the two phases. This kind of critical point is present in our model of vegetation-fire spreading as well, as we can show analytically in the mean-field case.

The contact process on the $2D$ lattice, however, displays another kind of phase transition related to its spatial structure \cite{martin2020intermittent}, a percolation transition. The order parameter of this transition is the probability that a site belongs to a spanning cluster, i.e., an infinite cluster, which is zero below the percolation transition and greater than zero above. Notice that Martín and collaborators \cite{martin2020intermittent} use a slightly different definition of the contact process, in which empty sites are occupied by neighbors with a probability $p$ and occupied sites become empty with a probability $1-p$. They show numerically that the percolation transition in a $2D$ lattice happens at $p_{\rm perc} \approx 0.725$. We can immediately recover our formulation by noting that $p = \lambda/(\lambda + \mu)$, giving the result used in the main text $(\lambda/\mu)_{\rm perc} \approx 2.63$.

\subsection*{Mean-field behavior of the model}
\noindent The mean-field equations read
\begin{align}
\label{eqn:MF}
    \dv{p_{\varnothing}}{t} & = - \lambda_V p_V p_{\varnothing} + d_Fp_F + d_Vp_V \nonumber \\
    \dv{p_F}{t} & = -d_F p_F + (\lambda_F p_F + b_F)p_V \\
    \dv{p_V}{t} & = -(d_V + b_F + \lambda_F p_F)p_V + \lambda_V p_V p_{\varnothing} \nonumber
\end{align}
where $p_{\varnothing}$, $p_F$ and $p_V$ are the probabilities of each state. Since $p_{\varnothing} = 1 - p_F - p_V$ we consider only the equations for $p_F$ and $p_V$. In general, the stationary state of the system is given by
\begin{equation}
    \begin{cases}
    d_F p_F = (\lambda_F p_F + b_F)p_V \\
    (d_V + b_F + \lambda_F p_F)p_V  =  \lambda_V p_V (1 - p_V - p_F)
    \end{cases}.
\end{equation}
These equations have an absorbing solution, since $(p^\text{abs}_V, p^\text{abs}_F) = (0,0)$ is a trivial solution of the system. Importantly, it is easy to show that by adding a birth term for the vegetation this empty solution disappears, as expected. The Jacobian matrix evaluated at $(p^\text{abs}_V, p^\text{abs}_F)$ is given by
\begin{equation}
    J^\text{abs} = \begin{pmatrix}-d_F & b_F \\ 0 & - (b_F + d_V) +\lambda_V \\\end{pmatrix}
\end{equation}
whose eigenvalues are $\mu_1^\text{abs} = -d_F$ and $\mu_2^\text{abs} = \lambda_V - b_F - d_V$. Thus, the empty state is only stable below $\lambda^\text{abs}_V = b_F + d_V$, which is the absorbing critical point of the system. Notice that $\lambda_F$ does not play a meaningful role in the stability of the empty state, a feature that is likely wrong in a spatially embedded model.

The other stationary state of the system is given by
\begin{align}
p_F^\text{stat} & = \frac{-d_V \lambda_F - (b_F + d_F + \lambda_F) \lambda_V + \sqrt{f^\text{stat}_{F, V}}}{2\lambda_F \lambda_V} \\
p_V^\text{stat} & = \frac{\lambda_F (-2b_F - d_V) -(b_F + d_F - \lambda_F)\lambda_V + \sqrt{f^\text{stat}_{F, V}}}{2\lambda_F (\lambda_F +\lambda_V)}
\end{align}
where $f^\text{stat}_{F, V} = 4d_F \lambda_V\lambda_F (\lambda^\text{abs}_V - \lambda_V) + (d_V \lambda_F - (b_F + d_F + \lambda_F) \lambda_V)^2$ is positive above $\lambda^\text{abs}_V$. The eigenvalues of $J^\text{st}$ always have a negative real part if $\lambda_V > b_F + d_V$ while they always may have a non-vanishing imaginary part. Hence, the relaxation towards the steady state typically happens in an oscillatory fashion. In particular, these oscillations play a major role in the evolution of the finite-size stochastic model, where the noise can push the system to the absorbing state or produce sustained stochastic oscillations, as we see in Figure~\ref{fig:model}i.

\subsection*{Exact stochastic simulation}
\noindent Simulations of the three-state model on a given network, such as a $2$-dimensional lattice, are performed using the Gillespie algorithm \cite{gillespie1977exact}. If we assume that there are $N$ sites in the network and $M$ possible transitions - in our model, $M = 6$ - then, at each time the network can be associated with a propensity matrix $A_{\mu i}^{(t)}$, where $\mu = 1, \dots, N$ and $i = 1, \dots, M$. Each row of $A_{\mu i}^{(t)}$ is given by the transition rates that the $\mu$-th site can undergo, given its state at time $t$. We introduce the total propensity $\alpha_0^{(t)} = \sum_\mu\sum_i A_{\mu i}^{(t)}$, so that the waiting time for the next transition is given by
\begin{equation}
    \tau^{(t)} = - \bigl(\alpha_0^{(t)}\bigl)^{-1}\log u
\end{equation}
where $u$ is uniformly distributed in $[0,1]$. Then, the transition $\bar{i}$ that occurs and the site $\bar{\mu}$ at which it occurs are such that
\begin{equation}
    \sum_{\mu = 1}^{\bar{\mu}-1}\sum_{i = 1}^{\bar{i}-1}A_{\mu i}^{(t)}\le \alpha_0^{(t)} v < \sum_{\mu = 1}^{\bar{\mu}}\sum_{i = 1}^{\bar{i}}A_{\mu i}^{(t)}
\end{equation}
where $v$ is once again uniformly distributed in $[0,1]$. We then update $A_{\bar{\mu}i}$ with the new transition rates for $\bar{\mu}$ and set the time to $t+\tau$.

\subsection*{Simulations and time-scale separation}
\noindent As pointed out in the main text, it is reasonable to expect that the vegetation dynamics is much slower than the fire dynamics. However, the parameter space of the model is extremely large and thus a phase space plot for the full model proves to be unfeasible. Therefore, in order to simplify the problem and reduce the number of free parameters, we assume that the vegetation configuration does not change during the propagation of a fire. This approximation is compatible with the time evolution of the model if we also assume that fires are rare events and the complete model predicts a charge-discharge dynamics. In fact, realistically, we expect the vegetation - if dry enough - to act as fuel during a fire propagation, which thus has to stop when locally all fuel is exhausted. Then the vegetation regrows and only after enough fuel is accumulated a new fire can start - that is, the two processes happen at different timescales.

This assumption is implemented in simulations as follows. Let us start with a network $\left(\{\sigma_i\}_{i=1}^N, \{E_{ij}\})\right)$ where the $N$ sites $\sigma_i$ are such that $\sigma_i \in \{\varnothing, V\}$ and $\{E_{ij}\}$ are the edges between the sites. We look for a stationary configuration $\{\sigma_i^\mathrm{stat}\}_{i=1}^N$ of the reactions
\begin{gather}
    (\sigma_i = V) + (\sigma_j = \varnothing) \xrightarrow{\lambda_V, E_{ij}} (\sigma_i = V) + (\sigma_j = V) \\
    (\sigma_i = V) \xrightarrow{d_V} (\sigma_i = \varnothing)
\end{gather}
where the notation $\xrightarrow{\lambda_V, E_{ij}}$ means that the reaction happens at a rate $\lambda_V$ if and only if $i$ and $j$ are joined by an edge $E_{ij}$. The system has an absorbing configuration $\{\sigma_i = \varnothing\}$ and its stationary configurations only depend on the ratio of the reaction rates $\zeta_V = \lambda_V/d_V$. See Supplementary Figures for examples of such configurations in a $2$-dimensional lattice.

Our approximation consists in obtaining a network over which the fires can spread from the stationary configuration $\{\sigma_i^\mathrm{stat}\}_{i=1}^N$. In particular, we consider the subgraph induced by the map $g : i \mapsto \mu$ defined for all the indexes $i$ such that $\sigma_i = V$. If we call these sites $s_\mu = \sigma_{g(i)}$, we end up with the vegetation subgraph $\left(\{s_\mu\}_{\mu=1}^{N_V}, \{E_{\mu\nu}\})\right)$ where $E_{\mu\nu} = E_{g(i)g(j)}$ and $N_V$ is the number of original vegetation sites. This subgraph is typically composed of many disjointed components. These components contain roughly the same number of nodes for $\zeta_V^\mathrm{abs} < \zeta_V \ll \zeta_V^\mathrm{perc}$, since the stationary configuration is dominated by a large number of small vegetation clusters, whereas as we approach $\zeta_V^\mathrm{perc}$ a giant component emerges and it is eventually dominant for $\zeta_V \gg \zeta_V^\mathrm{perc}$.

We now assume that $s_\mu \in \{\varnothing, V, F\}$, and notice that the initial configuration is such that $s_\mu = V$, $\forall \mu = 1, \dots, N_V$. In order to sample the distribution of the fire sizes we can choose a site $s_{\bar{\mu}}$ and set $s_{\bar{\mu}} = F$. Then, to simulate a fire, we consider the reactions
\begin{gather}
    (s_\mu = F) + (s_\nu = V) \xrightarrow{\lambda_F, E_{\mu\nu}} (s_\mu = F) + (s_\nu = F) \\
    (s_\mu = F) \xrightarrow{d_F} (s_\mu = \varnothing)
\end{gather}
until there are no more $F$ sites in the network. Thus, the fire dynamics only depend on $\zeta_F = d_F/\lambda_F$. The fire size - i.e., the burned area - is simply the number of empty sites $N_{\varnothing}$ of the final configuration.

One should be careful that if $s_{\bar{\mu}}$ is chosen at random between all sites we typically favor larger components of the vegetation subgraph. Thus, we first uniformly sample a given component $C_s$ of the vegetation subgraph, and then we randomly choose a site within the selected component and set $C_s \ni s_{\bar{\mu}} = F$. This assumption is qualitatively equivalent to the assumption that if two fires start in the same cluster they will contribute to the same burned area. To be precise, this is only true if $\zeta_F$ is small enough, so that two fires inside the same component will coalesce with high probability. However, for larger values of $\zeta_F$ we expect fires to be  small independently of the size of the underlying component, hence our assumption does not affect the results. In this way, we are now able to computationally explore the model's behavior effectively and systematically. In particular, for each value of $\zeta_V$, we simulate a large number of stationary configurations $\{\sigma_i^\mathrm{stat}\}_{i=1}^N$. Then, for each of these configurations, at a given value $\zeta_F$ we simulate a number of fires much larger than the number of components $C_s$, thus ending up with a set of burned areas $\{N_{\varnothing}\}$ that gives us the fire size distribution at $(\zeta_F, \zeta_V)$.

\newpage
\pagebreak

\setcounter{equation}{0}
\setcounter{figure}{0}
\setcounter{table}{0}
\setcounter{page}{1}
\setcounter{section}{0}
\setcounter{subsection}{0}
\makeatletter
\renewcommand{\theequation}{S\arabic{equation}}
\renewcommand{\thefigure}{S\arabic{figure}}
\renewcommand{\thesection}{S\Roman{section}} 

\section*{Supplementary Figures}

\begin{figure}[htbp!]
    \centering
    \includegraphics[width = 7cm]{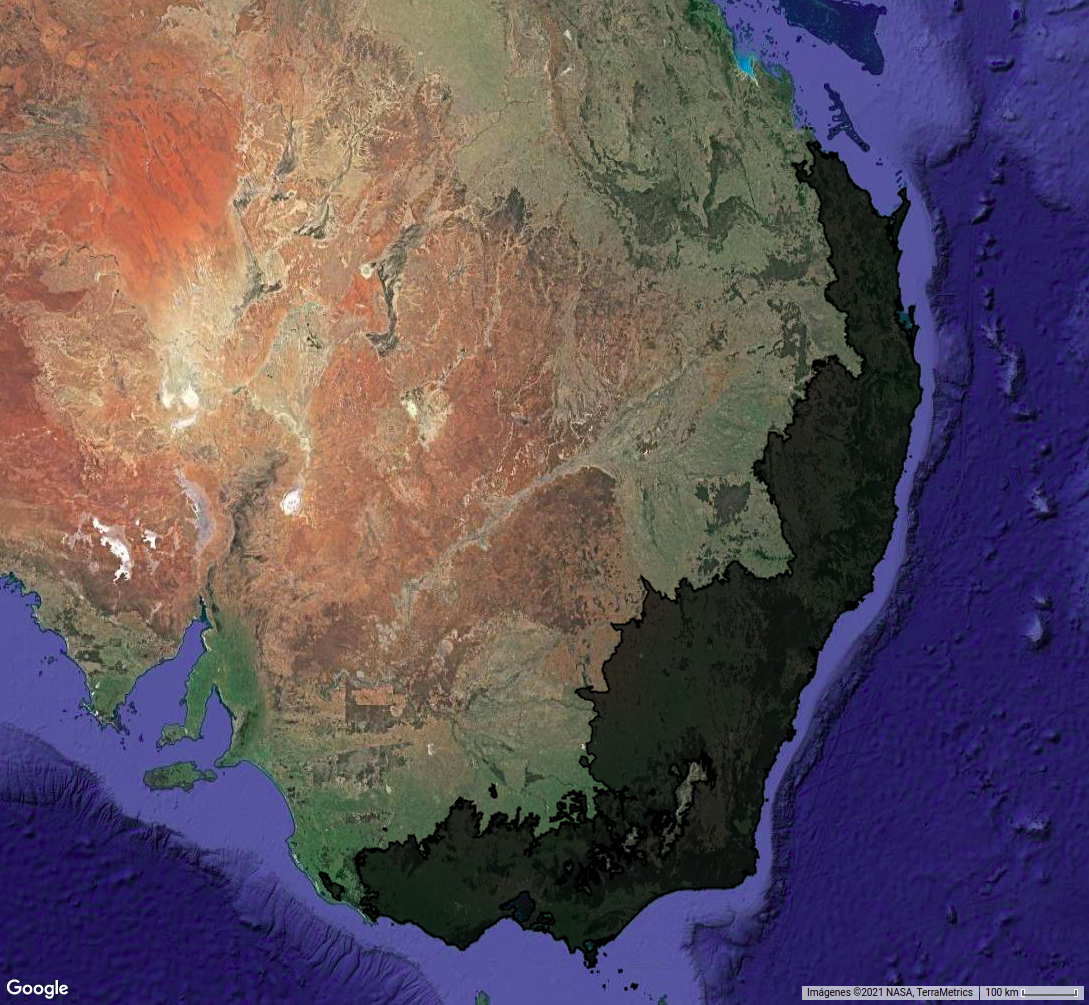}
\caption{\textbf{Region of study}. The shaded area represents the region of study encompassing the East and Southeast temperate broadleaf and mixed forests of continental Australia.}
\label{fig:australia}
\end{figure}

\vspace*{2cm}

\begin{figure}[htbp!]
\centering
    \includegraphics[width = 0.8\textwidth]{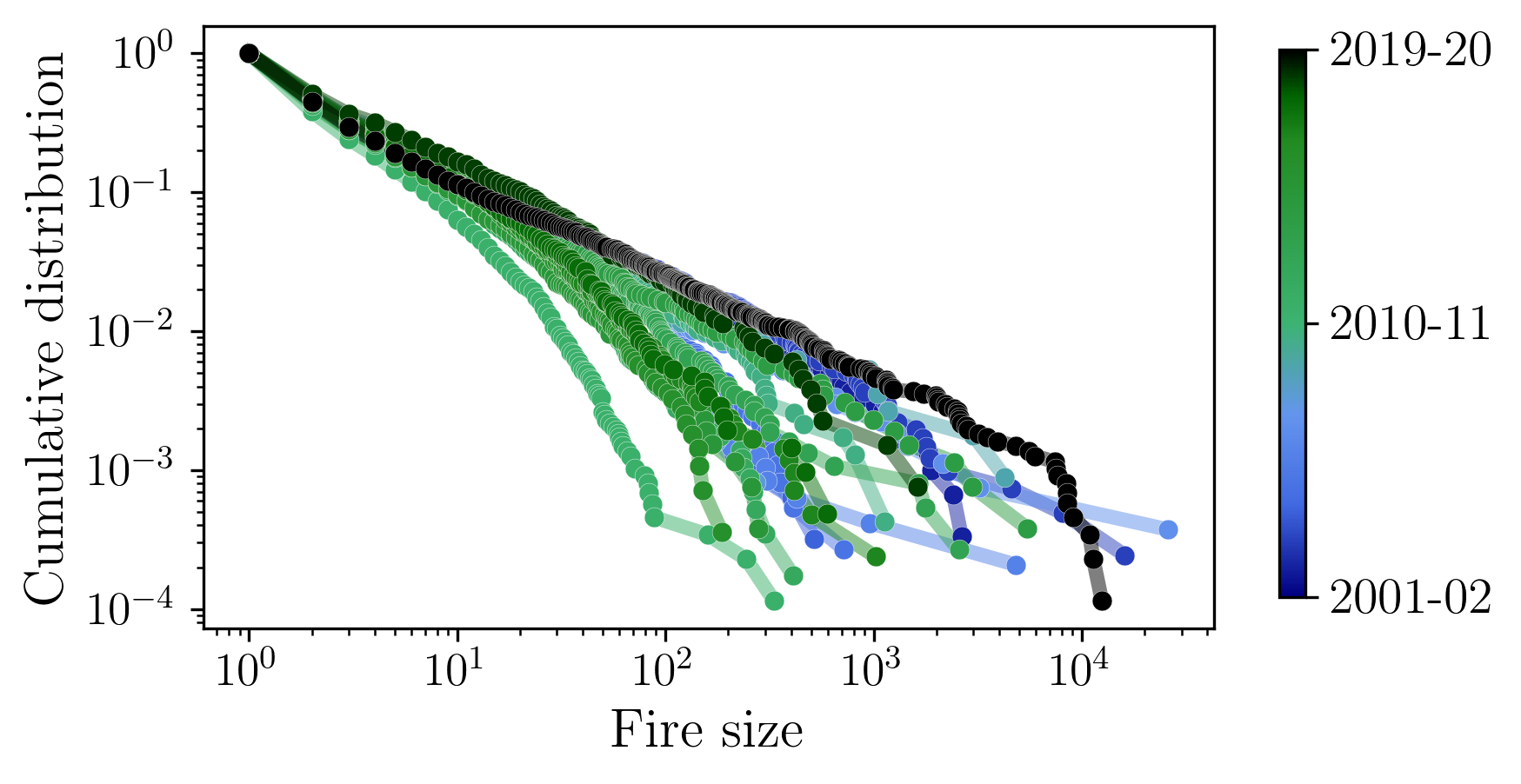}
\caption{\textbf{Comparison of yearly distributions of fire sizes from 2001-2002 to 2019-2002.}}
\label{fig:all_dist}
\end{figure}

\begin{figure}[htbp!]
    \centering
    \includegraphics[width = 1\textwidth]{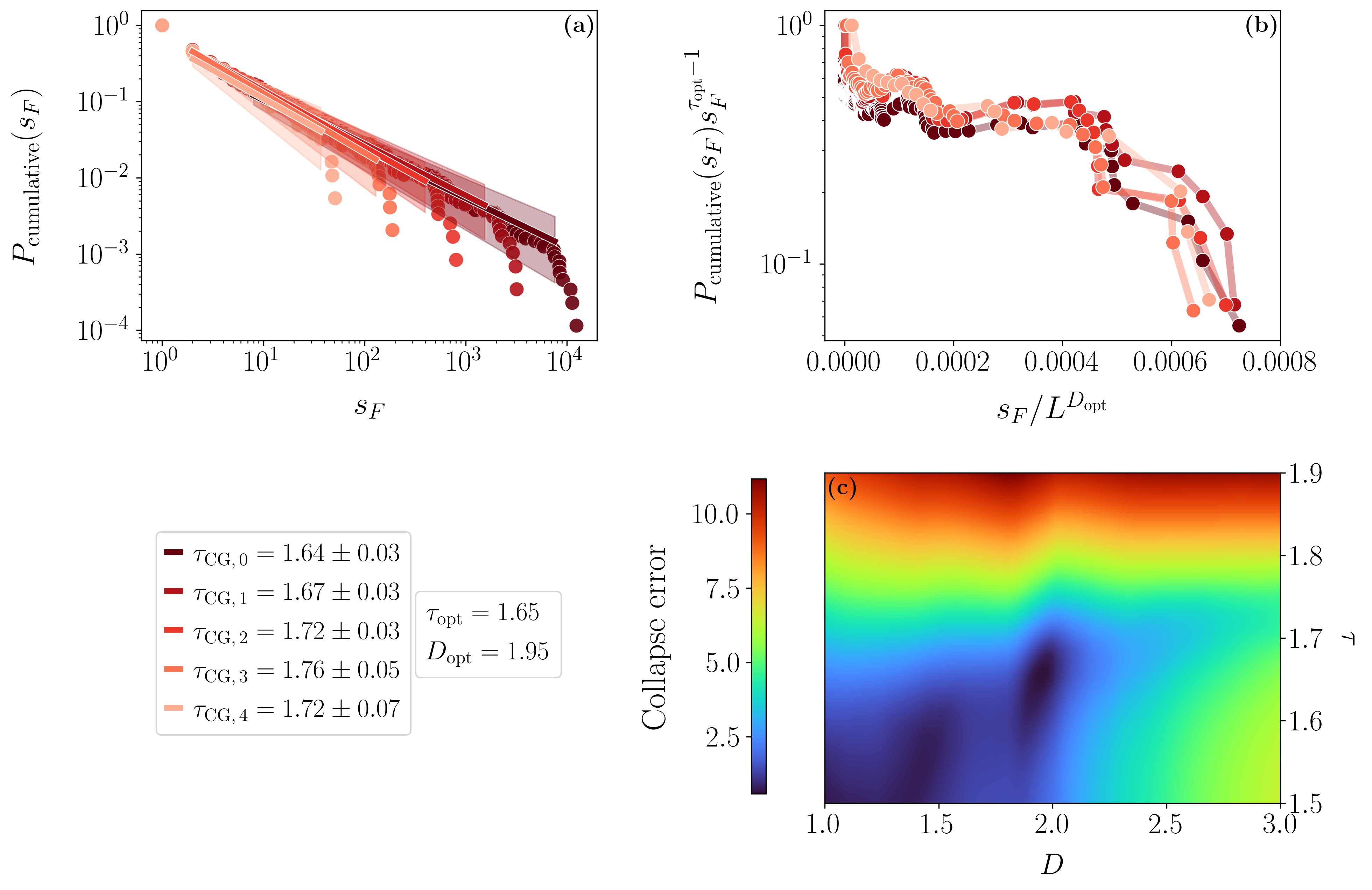}
\caption{\textbf{Invariance under coarse-graining and data collapse.} (a) The cumulative distribution of the fire sizes during 2019-2020 along the coarse-graining transformations, with the best power-law fit obtained by maximum-likelihood \cite{clauset2009power}. The exponents remain compatible at different CG steps. Notice that the different $\tau$ reported here those of the fire size distribution, not of the cumulative distribution. (b) The cumulative distributions of the data at different CG steps collapse into the same curve, once appropriately rescaled, as predicted by finite-size scaling. In order to appreciate the quality of the collapse, notice the different units in the vertical axis with respect to panel (a). (c) The two parameters $\tau$ and $D$ that are needed to collapse the cumulative distributions of the fire sizes can be chosen so that the collapse error will be minimal \cite{bhattacharjee2001measure}. As expected, we find $\tau_{\rm opt} \approx 1.65$, which is compatible with the fitted exponent.}
\label{fig:data_collapse}
\end{figure}

\vspace*{1cm}

\begin{figure}[htbp!]
\centering
    \includegraphics[width = 0.8\textwidth]{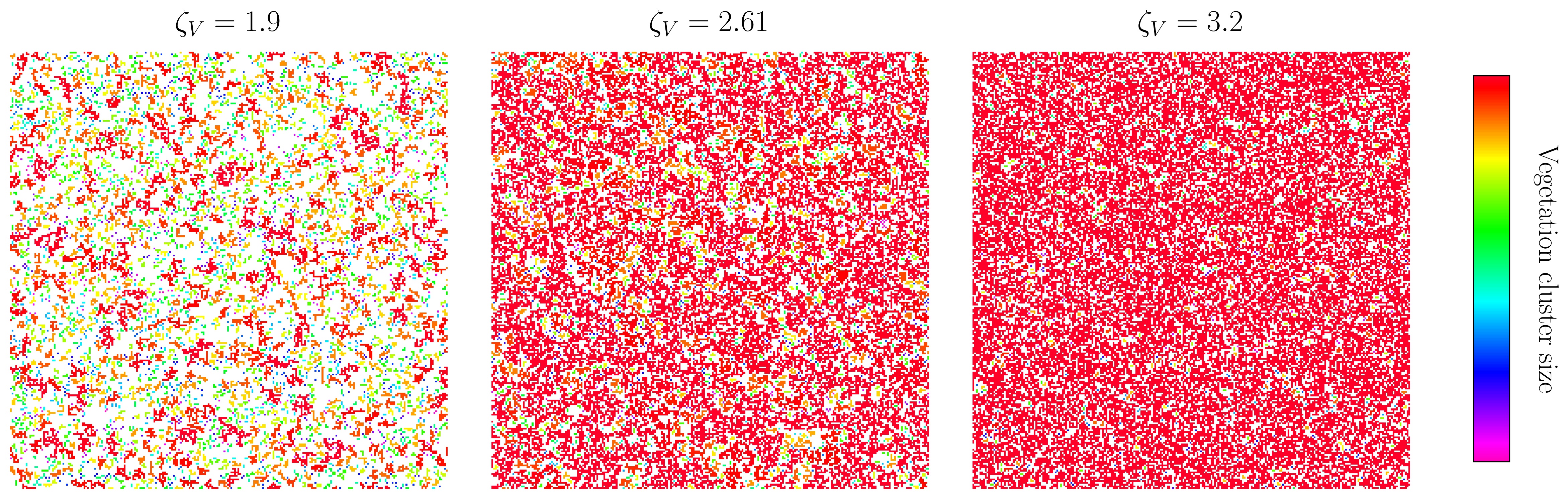}
\caption{\textbf{Stationary configurations of the vegetation contact process}. Different colors represent clusters of different sizes in a $250 \times 250$ $2$-dimensional lattice and t different values of $\zeta_V$. Notice how at the percolation threshold a system-size cluster appears.}
\label{fig:vegetation_conf}
\end{figure}


\clearpage

\begin{thebibliography}{10}

\bibitem{yates2008big}
C.~P. Yates, A.~C. Edwards, and J.~Russell-Smith, ``Big fires and their
  ecological impacts in australian savannas: size and frequency matters,'' {\em
  International Journal of Wildland Fire}, vol.~17, no.~6, pp.~768--781, 2008.

\bibitem{lindenmayer2020new}
D.~B. Lindenmayer and C.~Taylor, ``New spatial analyses of australian wildfires
  highlight the need for new fire, resource, and conservation policies,'' {\em
  Proceedings of the National Academy of Sciences}, vol.~117, no.~22,
  pp.~12481--12485, 2020.

\bibitem{deb2020causes}
P.~Deb, H.~Moradkhani, P.~Abbaszadeh, A.~S. Kiem, J.~Engstr{\"o}m,
  D.~Keellings, and A.~Sharma, ``Causes of the widespread 2019--2020 australian
  bushfire season,'' {\em Earth's Future}, vol.~8, no.~11, p.~e2020EF001671,
  2020.

\bibitem{ward2020impact}
M.~Ward, A.~I. Tulloch, J.~Q. Radford, B.~A. Williams, A.~E. Reside, S.~L.
  Macdonald, H.~J. Mayfield, M.~Maron, H.~P. Possingham, S.~J. Vine, {\em
  et~al.}, ``Impact of 2019--2020 mega-fires on australian fauna habitat,''
  {\em Nature Ecology \& Evolution}, vol.~4, no.~10, pp.~1321--1326, 2020.

\bibitem{phillips2020race}
N.~Phillips and B.~Nogrady, ``The race to decipher how climate change
  influenced australia's record fires,'' {\em Nature}, vol.~577, no.~7791,
  pp.~610--613, 2020.

\bibitem{dowdy2018climatological}
A.~J. Dowdy, ``Climatological variability of fire weather in australia,'' {\em
  Journal of Applied Meteorology and Climatology}, vol.~57, no.~2,
  pp.~221--234, 2018.

\bibitem{yu2020bushfires}
P.~Yu, R.~Xu, M.~J. Abramson, S.~Li, and Y.~Guo, ``Bushfires in australia: a
  serious health emergency under climate change,'' {\em The Lancet Planetary
  Health}, vol.~4, no.~1, pp.~e7--e8, 2020.

\bibitem{boer2020unprecedented}
M.~M. Boer, V.~R. de~Dios, and R.~A. Bradstock, ``Unprecedented burn area of
  australian mega forest fires,'' {\em Nature Climate Change}, vol.~10, no.~3,
  pp.~171--172, 2020.

\bibitem{arriagada2020climate}
N.~B. Arriagada, D.~M. Bowman, A.~J. Palmer, and F.~H. Johnston, ``Climate
  change, wildfires, heatwaves and health impacts in australia,'' in {\em
  Extreme Weather Events and Human Health}, pp.~99--116, Springer, 2020.

\bibitem{abram2021connections}
N.~J. Abram, B.~J. Henley, A.~Sen~Gupta, T.~J. Lippmann, H.~Clarke, A.~J.
  Dowdy, J.~J. Sharples, R.~H. Nolan, T.~Zhang, M.~J. Wooster, {\em et~al.},
  ``Connections of climate change and variability to large and extreme forest
  fires in southeast australia,'' {\em Communications Earth \& Environment},
  vol.~2, no.~1, pp.~1--17, 2021.

\bibitem{scheffer2009early}
M.~Scheffer, J.~Bascompte, W.~A. Brock, V.~Brovkin, S.~R. Carpenter, V.~Dakos,
  H.~Held, E.~H. Van~Nes, M.~Rietkerk, and G.~Sugihara, ``Early-warning signals
  for critical transitions,'' {\em Nature}, vol.~461, no.~7260, pp.~53--59,
  2009.

\bibitem{van2016you}
E.~H. van Nes, B.~M. Arani, A.~Staal, B.~van~der Bolt, B.~M. Flores,
  S.~Bathiany, and M.~Scheffer, ``What do you mean,‘tipping point’?,'' {\em
  Trends in ecology \& evolution}, vol.~31, no.~12, pp.~902--904, 2016.

\bibitem{rocha2018cascading}
J.~C. Rocha, G.~Peterson, {\"O}.~Bodin, and S.~Levin, ``Cascading regime shifts
  within and across scales,'' {\em Science}, vol.~362, no.~6421,
  pp.~1379--1383, 2018.

\bibitem{scheffer2020critical}
M.~Scheffer, {\em Critical transitions in nature and society}.
\newblock Princeton University Press, 2020.

\bibitem{wuyts2017amazonian}
B.~Wuyts, A.~R. Champneys, and J.~I. House, ``Amazonian forest-savanna
  bistability and human impact,'' {\em Nature Communications}, vol.~8, no.~1,
  pp.~1--12, 2017.

\bibitem{lovejoy2018amazon}
T.~E. Lovejoy and C.~Nobre, ``Amazon tipping point,'' {\em Science Advances},
  vol.~4, no.~2, p.~eaat2340, 2018.

\bibitem{scanlon2007positive}
T.~M. Scanlon, K.~K. Caylor, S.~A. Levin, and I.~Rodriguez-Iturbe, ``Positive
  feedbacks promote power-law clustering of kalahari vegetation,'' {\em
  Nature}, vol.~449, no.~7159, pp.~209--212, 2007.

\bibitem{taubert2018global}
F.~Taubert, R.~Fischer, J.~Groeneveld, S.~Lehmann, M.~S. M{\"u}ller,
  E.~R{\"o}dig, T.~Wiegand, and A.~Huth, ``Global patterns of tropical forest
  fragmentation,'' {\em Nature}, vol.~554, no.~7693, pp.~519--522, 2018.

\bibitem{Saravia2018a}
L.~A. Saravia, S.~Doyle, and B.~Bond-Lamberty, ``Power laws and critical
  fragmentation in global forests,'' {\em Scientific Reports}, vol.~8,
  p.~17766, 2018.

\bibitem{binney1992theory}
J.~J. Binney, N.~J. Dowrick, A.~J. Fisher, and M.~E. Newman, {\em The theory of
  critical phenomena: an introduction to the renormalization group}.
\newblock Oxford University Press, 1992.

\bibitem{caldarelli2001}
G.~Caldarelli, R.~Frondoni, A.~Gabrielli, M.~Montuori, R.~Retzlaff, and
  C.~Ricotta, ``Percolation in real wildfires,'' {\em Europhysics Letters},
  vol.~56, pp.~510--516, nov 2001.

\bibitem{sornette2006critical}
D.~Sornette, {\em Critical phenomena in natural sciences: chaos, fractals,
  selforganization and disorder: concepts and tools}.
\newblock Springer Science \& Business Media, 2006.

\bibitem{mckenzie2012power}
D.~McKenzie and M.~C. Kennedy, ``Power laws reveal phase transitions in
  landscape controls of fire regimes,'' {\em Nature Communications}, vol.~3,
  no.~1, pp.~1--6, 2012.

\bibitem{Dinerstein2017}
E.~Dinerstein, D.~Olson, A.~Joshi, C.~Vynne, N.~D. Burgess, E.~Wikramanayake,
  N.~Hahn, S.~Palminteri, P.~Hedao, R.~Noss, M.~Hansen, H.~Locke, E.~C. Ellis,
  B.~Jones, C.~V. Barber, R.~Hayes, C.~Kormos, V.~Martin, E.~Crist,
  W.~Sechrest, L.~Price, J.~E.~M. Baillie, D.~Weeden, K.~Suckling, C.~Davis,
  N.~Sizer, R.~Moore, D.~Thau, T.~Birch, P.~Potapov, S.~Turubanova,
  A.~Tyukavina, N.~{de Souza}, L.~Pintea, J.~C. Brito, O.~A. Llewellyn, A.~G.
  Miller, A.~Patzelt, S.~A. Ghazanfar, J.~Timberlake, H.~Kl{\"o}ser,
  Y.~{Shennan-Farp{\'o}n}, R.~Kindt, J.-P.~B. Lilles{\o}, P.~{van Breugel},
  L.~Graudal, M.~Voge, K.~F. {Al-Shammari}, and M.~Saleem, ``An
  {{Ecoregion}}-{{Based Approach}} to {{Protecting Half}} the {{Terrestrial
  Realm}},'' {\em BioScience}, vol.~67, pp.~534--545, June 2017.

\bibitem{Giglio2016}
L.~Giglio, W.~Schroeder, and C.~O. Justice, ``The collection 6 {{MODIS}} active
  fire detection algorithm and fire products,'' {\em Remote Sensing of
  Environment}, vol.~178, pp.~31--41, 2016.

\bibitem{stauffer2018introduction}
D.~Stauffer and A.~Aharony, {\em Introduction to percolation theory}.
\newblock CRC press, 2018.

\bibitem{grassberger1991forestfires}
P.~Grassberger and H.~Kantz, ``On a forest fire model with supposed
  self-organized criticality,'' {\em Journal of Statistical Physics}, vol.~63,
  no.~3, pp.~685--700, 1991.

\bibitem{drossel1992self}
B.~Drossel and F.~Schwabl, ``Self-organized critical forest-fire model,'' {\em
  Physical review letters}, vol.~69, no.~11, p.~1629, 1992.

\bibitem{grassberger1993forestfires}
P.~Grassberger, ``On a self-organized critical forest-fire model,'' {\em
  Journal of Physics A: Mathematical and General}, vol.~26, no.~9, p.~2081,
  1993.

\bibitem{Christensen1993forestfires}
K.~Christensen, H.~Flyvbjerg, and Z.~Olami, ``Self-organized critical
  forest-fire model: Mean-field theory and simulation results in 1 to 6
  dimenisons,'' {\em Phys. Rev. Lett.}, vol.~71, pp.~2737--2740, Oct 1993.

\bibitem{malamud1998forest}
B.~D. Malamud, G.~Morein, and D.~L. Turcotte, ``Forest fires: an example of
  self-organized critical behavior,'' {\em Science}, vol.~281, no.~5384,
  pp.~1840--1842, 1998.

\bibitem{Turcotte2002forestfires}
D.~L. Turcotte, B.~D. Malamud, F.~Guzzetti, and P.~Reichenbach,
  ``Self-organization, the cascade model, and natural hazards,'' {\em
  Proceedings of the National Academy of Sciences}, vol.~99, no.~suppl\_1,
  pp.~2530--2537, 2002.

\bibitem{palmieri2020forest}
L.~Palmieri and H.~J. Jensen, ``The forest fire model: the subtleties of
  criticality and scale invariance,'' {\em Frontiers in Physics}, vol.~8,
  p.~257, 2020.

\bibitem{dickman1999}
J.~Marro and R.~Dickman, {\em Nonequilibrium Phase Transitions in Lattice
  Models}.
\newblock Cambridge University Press, 1999.

\bibitem{henkel2009}
M.~Henkel, H.~Hinrichsen, and S.~L{\"u}beck, {\em Non-Equilibrium Phase
  Transitions. Volume 1: Absorbing Phase Transitions}.
\newblock Springer, 2009.

\bibitem{clauset2009power}
A.~Clauset, C.~R. Shalizi, and M.~E. Newman, ``Power-law distributions in
  empirical data,'' {\em SIAM review}, vol.~51, no.~4, pp.~661--703, 2009.

\bibitem{marsili2013sampling}
M.~Marsili, I.~Mastromatteo, and Y.~Roudi, ``On sampling and modeling complex
  systems,'' {\em Journal of Statistical Mechanics: Theory and Experiment},
  vol.~2013, no.~09, p.~P09003, 2013.

\bibitem{gerlach2019testing}
M.~Gerlach and E.~G. Altmann, ``Testing statistical laws in complex systems,''
  {\em Physical review letters}, vol.~122, no.~16, p.~168301, 2019.

\bibitem{serafino2021true}
M.~Serafino, G.~Cimini, A.~Maritan, A.~Rinaldo, S.~Suweis, J.~R. Banavar, and
  G.~Caldarelli, ``True scale-free networks hidden by finite size effects,''
  {\em Proceedings of the National Academy of Sciences}, vol.~118, no.~2,
  p.~e2013825118, 2021.

\bibitem{pikovsky2002synchronization}
A.~Pikovsky, M.~Rosenblum, and J.~Kurths, ``Synchronization: a universal
  concept in nonlinear science,'' 2002.

\bibitem{wilson1983renormalization}
K.~G. Wilson, ``The renormalization group and critical phenomena,'' {\em
  Reviews of Modern Physics}, vol.~55, no.~3, p.~583, 1983.

\bibitem{loreto1995renormalization}
V.~Loreto, L.~Pietronero, A.~Vespignani, and S.~Zapperi, ``Renormalization
  group approach to the critical behavior of the forest-fire model,'' {\em
  Physical review letters}, vol.~75, no.~3, p.~465, 1995.

\bibitem{goldenfeld2018lectures}
N.~Goldenfeld, {\em Lectures on phase transitions and the renormalization
  group}.
\newblock CRC Press, 2018.

\bibitem{jona1975renormalization}
G.~Jona-Lasinio, ``The renormalization group: A probabilistic view,'' {\em Il
  Nuovo Cimento B (1971-1996)}, vol.~26, no.~1, pp.~99--119, 1975.

\bibitem{bak1990forest}
P.~Bak, K.~Chen, and C.~Tang, ``A forest-fire model and some thoughts on
  turbulence,'' {\em Physics letters A}, vol.~147, no.~5-6, pp.~297--300, 1990.

\bibitem{bak2013nature}
P.~Bak, {\em How nature works: the science of self-organized criticality}.
\newblock Copernicus Books, New York, 1996.

\bibitem{PhysRevE.104.L012201}
D.~Rybski, V.~Butsic, and J.~W. Kantelhardt, ``Self-organized multistability in
  the forest fire model,'' {\em Phys. Rev. E}, vol.~104, p.~L012201, Jul 2021.

\bibitem{pueyo2007self}
S.~Pueyo, ``Self-organised criticality and the response of wildland fires to
  climate change,'' {\em Climatic Change}, vol.~82, no.~1, pp.~131--161, 2007.

\bibitem{staal2018resilience}
A.~Staal, E.~H. van Nes, S.~Hantson, M.~Holmgren, S.~C. Dekker, S.~Pueyo,
  C.~Xu, and M.~Scheffer, ``Resilience of tropical tree cover: The roles of
  climate, fire, and herbivory,'' {\em Global Change Biology}, vol.~24, no.~11,
  pp.~5096--5109, 2018.

\bibitem{zinck2011shifts}
R.~D. Zinck, M.~Pascual, and V.~Grimm, ``Understanding shifts in wildfire
  regimes as emergent threshold phenomena.,'' {\em The American Naturalist},
  vol.~178, no.~6, pp.~E149--E161, 2011.
\newblock PMID: 22089877.

\bibitem{de2015structural}
M.~De~Domenico, V.~Nicosia, A.~Arenas, and V.~Latora, ``Structural reducibility
  of multilayer networks,'' {\em Nature communications}, vol.~6, no.~1,
  pp.~1--9, 2015.

\bibitem{de2016physics}
M.~De~Domenico, C.~Granell, M.~A. Porter, and A.~Arenas, ``The physics of
  spreading processes in multilayer networks,'' {\em Nature Physics}, vol.~12,
  no.~10, pp.~901--906, 2016.

\bibitem{kivela2014multilayer}
M.~Kivel{\"a}, A.~Arenas, M.~Barthelemy, J.~P. Gleeson, Y.~Moreno, and M.~A.
  Porter, ``Multilayer networks,'' {\em Journal of complex networks}, vol.~2,
  no.~3, pp.~203--271, 2014.

\bibitem{harris1974}
T.~E. Harris, ``Contact interactions on a lattice,'' {\em Annals of
  Probability}, vol.~2, no.~6, 1974.

\bibitem{marro2005nonequilibrium}
J.~Marro and R.~Dickman, {\em Nonequilibrium phase transitions in lattice
  models}.
\newblock Cambridge University Press, 2005.

\bibitem{gillespie1977exact}
D.~T. Gillespie, ``Exact stochastic simulation of coupled chemical reactions,''
  {\em The journal of physical chemistry}, vol.~81, no.~25, pp.~2340--2361,
  1977.

\bibitem{martin2020intermittent}
P.~V. Mart{\'\i}n, V.~Dom{\'\i}nguez-Garc{\'\i}a, and M.~A. Mu{\~n}oz,
  ``Intermittent percolation and the scale-free distribution of vegetation
  clusters,'' {\em New Journal of Physics}, vol.~22, no.~8, p.~083014, 2020.

\bibitem{Keeley2011}
J.~E. Keeley, J.~G. Pausas, P.~W. Rundel, W.~J. Bond, and R.~A. Bradstock,
  ``Fire as an evolutionary pressure shaping plant traits,'' {\em Trends in
  Plant Science}, vol.~16, pp.~406--411, Aug. 2011.

\bibitem{jyoteeshkumar2021intensifying}
P.~Jyoteeshkumar~reddy, S.~E. Perkins-Kirkpatrick, and J.~J. Sharples,
  ``Intensifying australian heatwave trends and their sensitivity to
  observational data,'' {\em Earth's Future}, vol.~9, no.~4, p.~e2020EF001924,
  2021.

\bibitem{Fairman2015}
T.~A. Fairman, C.~R. Nitschke, L.~T. Bennett, T.~A. Fairman, C.~R. Nitschke,
  and L.~T. Bennett, ``Too much, too soon? {{A}} review of the effects of
  increasing wildfire frequency on tree mortality and regeneration in temperate
  eucalypt forests,'' {\em International Journal of Wildland Fire}, vol.~25,
  pp.~831--848, Sept. 2015.

\bibitem{steel2021ecological}
Z.~L. Steel, D.~Foster, M.~Coppoletta, J.~M. Lydersen, S.~L. Stephens,
  A.~Paudel, S.~H. Markwith, K.~Merriam, and B.~M. Collins, ``Ecological
  resilience and vegetation transition in the face of two successive large
  wildfires,'' {\em Journal of Ecology}, 2021.

\bibitem{nolan2021limits}
R.~H. Nolan, L.~Collins, A.~Leigh, M.~K. Ooi, T.~J. Curran, T.~A. Fairman,
  V.~R. de~Dios, and R.~Bradstock, ``Limits to post-fire vegetation recovery
  under climate change,'' {\em Plant, cell \& environment}, 2021.

\bibitem{de2013fire}
V.~de~L.~Dantas, M.~A. Batalha, and J.~G. Pausas, ``Fire drives functional
  thresholds on the savanna--forest transition,'' {\em Ecology}, vol.~94,
  no.~11, pp.~2454--2463, 2013.

\bibitem{oliveras2016many}
I.~Oliveras and Y.~Malhi, ``Many shades of green: the dynamic tropical
  forest--savannah transition zones,'' {\em Philosophical Transactions of the
  Royal Society B: Biological Sciences}, vol.~371, no.~1703, p.~20150308, 2016.

\bibitem{cranmer2020frontier}
K.~Cranmer, J.~Brehmer, and G.~Louppe, ``The frontier of simulation-based
  inference,'' {\em Proceedings of the National Academy of Sciences}, vol.~117,
  no.~48, pp.~30055--30062, 2020.

\bibitem{nicoletti2021mutual}
G.~Nicoletti and D.~M. Busiello, ``Mutual information disentangles interactions
  from changing environments,'' {\em Physical Review Letters}, vol.~127,
  no.~22, p.~228301, 2021.

\bibitem{nicoletti2020scaling}
G.~Nicoletti, S.~Suweis, and A.~Maritan, ``Scaling and criticality in a
  phenomenological renormalization group,'' {\em Phys. Rev. Research}, vol.~2,
  p.~023144, May 2020.

\bibitem{giorgio_nicoletti_2023_7540390}
G.~Nicoletti, ``{Stochastic simulation of a spatial forest-fire model},'' {\em
  Zenodo}, Jan. 2023.

\bibitem{lsaravia_2023_7541674}
L.~Saravia, ``{Code for download and Data for Burned Area Australia},'' {\em
  Zenodo}, Jan. 2023.

\bibitem{RStat2021}
{R Core Team}, {\em R: A Language and Environment for Statistical Computing}.
\newblock R Foundation for Statistical Computing, Vienna, Austria, 2021.

\bibitem{bhattacharjee2001measure}
S.~M. Bhattacharjee and F.~Seno, ``A measure of data collapse for scaling,''
  {\em Journal of Physics A: Mathematical and General}, vol.~34, no.~33,
  p.~6375, 2001.

\end{thebibliography}
\end{document}